\documentclass[12pt]{article}
\usepackage{amssymb,amsmath,epsfig}
\usepackage{graphicx}
\allowdisplaybreaks
\begin{document}
\title{\textbf{Feasible Stellar Interiors Beyond Einstein Gravity:
Insights from Non-Metricity-Matter Coupled Gravitational Theory}}
\author{M. Sharif$^{1,2}$\thanks{msharif.math@pu.edu.pk} ,
M. Zeeshan Gul$^{3,4,1}$\thanks{mzeeshangul.math@gmail.com}
~and Adeeba Arooj$^{1}$\thanks{aarooj933@gmail.com}\\
$^1$ Department of Mathematics and Statistics, The University of Lahore,\\
1-KM Defence Road Lahore-54000, Pakistan.\\
$^2$ Research Center of Astrophysics and Cosmology, Khazar
University,\\ Baku, AZ1096, 41 Mehseti Street, Azerbaijan. \\
$^3$ College of Transportation, Tongji University, Shanghai 201804,
China.\\
$^4$ Postdoctoral Station of Mechanical Engineering, Tongji,
University, \\ Shanghai 201804, China.}

\date{}
\maketitle

\begin{abstract}
This manuscript examines viability and stability of anisotropic
compact objects in the framework of $f(Q,L_m)$ gravity ($Q$ is the
non-metricity and $L_m$ is the matter Lagrangian). We assume a
particular functional form of this theory to get explicit
expressions for the field equations which govern the behavior of
matter and geometry in this context. The configuration of static
spherically symmetric structures is evaluated using the two
innovative non-singular solutions. We use smooth matching conditions
to evaluate the values of unknown constants in the metric
coefficients. The viability of considered compact stars is assessed
using a graphic analysis of various important physical
characteristics. We also investigate stability of the considered
stellar objects through sound speed method. It is found that these
stellar objects are viable and stable, as all the required
conditions are satisfied.
\end{abstract}
\textbf{Keywords:} Stellar interiors; Energy conditions;
Alternative gravitational framework; Stability analysis.\\
\textbf{PACS:} 97.60.Lf; 97.10.Fy; 04.50.Kd; 98.58.M.

\section{Introduction}

Einstein's gravitational theory (GR) provides a geometric framework
that remains foundational pillar of modern physics. This theory
faces a problem when it explains the phenomenon of late-time cosmic
expansion. There are many astronomical observations, including
supernovae and large structures which indicate that the universe is
in an expansion phase \cite{1}. This acceleration is hypothesized to
originate from an ambiguous force, named as dark energy (DE). The
enigmatic nature of DE constitutes unresolved questions in
cosmology. To address this issue, the $\Lambda$CDM model has been
established to describe the ambiguous aspects of DE. While
$\Lambda$CDM model aligns well with observational constraints, but
it faces theoretical limitations \cite{2}-\cite{6}. These challenges
inspired the scientific community to modify GR to gain deep insights
for DE and cosmic acceleration \cite{7}. Such modifications redefine
the geometric components of the Einstein-Hilbert action, thereby
proposing alternative frameworks to resolve mysteries of the
universe. General theory of relativity can be conceptualized through
two primary frameworks: one based on curvature without torsion and
non-metricity and another based on teleparallelism with zero
curvature and non-metricity. The non-metricity describes changes in
vector's length at the time of parallel transport where the non-zero
covariant divergence of metric tensor exists, leading to symmetric
teleparallel theory \cite{12}. A comprehensive examination of this
theory can be found in \cite{z1}-\cite{z8}. Different theoretical
models and the limitations have been thoroughly investigated in
\cite{MT1}-\cite{13o}.

Hazarika et al \cite{34} extended the symmetric teleparallel theory
by incorporating the matter-Lagrangian into the functional action,
known as $f(Q,L_{m})$ theory. This gravitational theory introduces a
novel coupling between geometric quantities and matter fields,
attracting a significant attention of the scientific community due
to its potential to understand gravitational phenomena. The
motivation of this theory arises from the need to comprehend its
theoretical implications and its role in astrophysics and cosmology.
Introducing non-metricity into the action expands the scope of
possible gravitational dynamics and allow for a wide variety of
viable cosmological solutions. Furthermore, the use of the
matter-Lagrangian enables this theory to capture how matter contents
influence the gravitational field. This modified theory seeks a
comprehensive approach to gravitational physics, addressing
observational challenges and enhancing our understanding of gravity
and the universe. Samaddar \cite{35} studied gravitational
baryogenesis in this modified framework. Gul et al \cite{36a}
examined the detailed investigation of the ghost DE model to assess
the influence of correction terms on cosmic evolution in this
modified framework. Myrzakulov et al examined cosmic expansion in
this theory \cite{36b}-\cite{36bbbb}.

The $f(Q,L_{m})$ theory introduces a specific connection of geometry
and matter, capturing considerable attention among researchers due
to its profound implications in gravitational physics. The
exploration of this theory is motivated by a desire to understand
its theoretical implications and significance in astrophysical as
well as cosmological contexts. In $f(Q,L_{m})$ theory, the inclusion
of non-metricity and the presence of matter source enable a more
detailed depiction of gravitational interactions. Non-metricity
represents the deviation from the Levi-Civita connection, which is
the connection compatible with the metric tensor in GR. By
incorporating non-metricity into the gravitational action, this
theory introduces additional degrees of freedom, leading to new
gravitational dynamics and cosmological solutions. Furthermore, the
dependence on the matter-Lagrangian allows this theory to capture
the effects of matter content on the gravitational field. The
recently discovered $f(Q,L_{m})$ theory highlight the diverse
avenues of research in reshaping our understanding of cosmic
evolution. Myrzakulov et al investigated cosmology within the
framework of modified gravity, while using the non-linear model
\cite{74}. Rana et al investigated the influence of viscosity on the
evolution of the cosmos within the framework of $f(Q,L_{m})$ gravity
\cite{75}. Sharif and Arooj studied stable stellar stars within the
modified framework \cite{75x}. Thus, the motivation behind this
theory is to consider a more comprehensive framework for
gravitational physics and cosmology, capable of addressing current
observational discrepancies and offering insights into fundamental
questions about the nature of gravity and the universe.

Astrophysical phenomena refer to the study of physical process that
occur in the universe. These phenomena encompass a wide range of
events from a behavior of stars and galaxies to the evolution of the
universe. Stars are important part of cosmos and occupy a state of
equilibrium until pressure from the nuclear fusion reactions
balances the gravitational pull. However, when a star collapses it
gives rise to a new dense objects, named as compact stars (CSs). In
astrophysics, the existence of CSs motivated a number of researchers
to examine their various evolutionary phases and internal
characteristics. Additionally, CSs offer important new information
about basic physics, including how matter behaves at extremely high
densities and how strong gravitational fields work
\cite{14}-\cite{25}. The study of stellar interiors in modified
theories has gained significant attention from researchers in recent
years. Maurya et al \cite{25aa} studied the interiors of ultra-dense
compact spherical systems in $f(R,\mathcal{T})$ theory ($R$ is Ricci
scalar and $\mathcal{T}$ defines the energy-momentum tensor). Maurya
and Tello-Ortiz \cite{26bb} extended this work by incorporating the
charge in fluid configuration and found that the charge acts as a
repulsive force, enhancing the stability of the system. Bamba et al
\cite{27cc} explored the effects of Brans-Dicke parameter and a
scalar field on the structural aspects of anisotropic spherical
spacetime. Jasim et al \cite{28dd} used gravitational decoupling
method to examine the anisotropic spherical solutions in $f(Q)$
theory. Sohail et al \cite{29ee} used non-singular constraint to
investigate a static anisotropic solution for compact stellar
objects in teleparallel gravity. Albalahi et al \cite{30ff} examined
the role of decoupling parameter on the charged compact spherical
structures. Naseer et al \cite{31gg} formulated three distinct
analytical solutions in Rastall theory. Pradhan et al \cite{32hh}
used Karmarkar condition to explore the deformed charged CSs in
$f(Q,\mathcal{T})$ theory. Maurya et al \cite{33ii} investigated
anisotropic stars with vanishing complexity constraint in the
framework of symmetric teleparallel theory.

We are motivated by aforementioned research to investigate unique
features of stellar interiors in $f(Q,L_{m})$ theory. Two distinct
non-singular solutions are studied in this framework. This paper
uses the following format. The theoretical basis for our approach is
established in section \textbf{2} where we derive the field
equations of this theory. The precise equations for matter contents
are then derived by certain model of this theory. A detailed
mathematical derivations and physical interpretations of both
solutions are provided in section \textbf{3}. Using matching
criteria, we assess the unknown characteristics of the solutions and
guarantee consistency between the exterior and interior solutions.
To determine feasibility of the CSs, we examine several physical
features in section \textbf{4}. In section \textbf{5}, we discuss
the stability of CSs. The last section summarizes our findings.

\section{Field Equations in $f(Q,L_{m})$ Gravity}

The mathematical basis of GR relies on Riemannian geometry, where
parallel transport preserves both direction and length of a vector
along a closed path. Weyl \cite{34a} presented a generalization in
which a vector's magnitude and orientation may vary during parallel
transport. This introduces a novel vector field $(Y^{\xi})$, which
determines the geometric structure of Weyl spacetime. In Weyl's
theory, the length of a vector changes as $\delta\ell= \ell Y_{\xi}
\delta x^ {\xi}$, where $\ell$ is the size of vector and $\delta
x^{\xi}$ defines infinitesimal path \cite{34a}. This implies that
the change in vector's length depends on connection coefficients,
initial length and displacement path. When the vector is transported
around a small closed surface enclosing an area $\delta
h^{\xi\tau}$, the resulting variation in its length is described by
$\delta\ell= \ell\Upsilon_{\xi\tau}\delta h^{\xi\tau}$, where
$\Upsilon_{\xi\tau}$ encapsulates the curvature associated with the
length variation
\begin{equation}\label{1a}
\Upsilon_{\xi\tau}=\nabla_{\tau}Y_{\xi}-\nabla_{\xi}Y_{\tau}.
\end{equation}
A spatial transformation $\hat{\ell}=\upsilon(x)\ell$ alters the
field equation $\hat{Y}_{\xi}$ to $\hat{Y}_{\xi}=Y_{\xi}+
(\ln\upsilon)_{\xi}$, while conformal transformations adjust the
components of the metric tensor as
$\hat{g}_{\xi\tau}=\upsilon^{2}g_{\xi\tau}$ and $\hat{g}^{\xi\tau}=
\upsilon^{-2}g^{\xi\tau}$. Weyl geometry encompasses various
characteristics, defined as
\begin{equation}\label{2a}
{\hat{\Gamma}}^{\vartheta}_{~\xi\tau}=\Gamma^{\vartheta}_{~\xi\tau}
+g_{\xi\tau}Y^{\vartheta}-\delta^{\vartheta}_{~\xi}Y_{\tau}-
\delta^{\vartheta}_{~\tau}Y_{\xi}.
\end{equation}
The Christoffel symbol $(\Gamma^{\vartheta}_{~\xi\tau})$ is the
affine connection in this framework. By requiring symmetry in the
modified connection coefficient
$(\hat{\Gamma}^{\vartheta}_{~\xi\tau})$, we can properly define a
gauge-covariant derivative. This derivative operator yields the Weyl
curvature tensor through standard geometric construction as
\begin{equation}\label{3a}
\hat{S}_{\xi\tau\vartheta\gamma}=\hat{S}
_{(\xi\tau)\vartheta\gamma}+\hat{S}_{[\xi\tau]\vartheta\gamma},
\end{equation}
where
\begin{equation}\nonumber
\hat{S}_{[\xi\tau]\vartheta\gamma}=S
_{\xi\tau\vartheta\gamma}+2\nabla_{\vartheta}Y_{[\xi
g_{\tau}]\gamma}+2\nabla_{\gamma}Y_{[\tau
g_{\xi}]\vartheta}+2Y_{\vartheta}Y_{[\xi
g_{\tau}]\vartheta}+2Y_{\gamma}Y_{[\tau
g_{\xi}]\vartheta}-2Y^{2}g_{\vartheta[\xi g_{\tau}]\gamma}.
\end{equation}
Applying the first contraction operation to the Weyl tensor, we have
\begin{eqnarray}\label{5a}
\hat{S}^{\xi}_{~\tau}&=&S^{\xi}_{~\tau}
+2Y^{\xi}Y_{\tau}+3\nabla_{\tau}Y^{\xi}-\nabla_{\xi}Y^{\tau}
+g^{\xi}_{~\tau}(\nabla_{\vartheta}Y^{\vartheta}
-2Y_{\vartheta}Y^{\vartheta}),\\\label{6a}
\hat{S}&=&\bar{S}^{\vartheta}_{~\vartheta}=
S+6(\nabla_{\xi}Y^{\xi}-Y_{\xi}Y^{\xi}).
\end{eqnarray}

Weyl-Cartan (WC) geometries extend beyond the Riemannian and Weyl
frameworks by incorporating non-zero torsion. The WC geometry uses a
symmetric metric tensor for length measurement and an asymmetric
connection that regulates parallel transport as
$d\varsigma^{\xi}=-\varsigma^{\vartheta}\hat{\Gamma}^{\xi}_{~\vartheta\tau}dx^{\tau}$.
The connection $(\tilde{\Gamma}^{\vartheta}_{~\xi\tau})$,
disformation tensor $(\Omega^{\vartheta}_{~\xi\tau})$ and contortion
tensor $(\Psi^{\vartheta}_{~\xi\tau})$ in this framework are given
as
\begin{eqnarray}\label{7a}
\tilde{\Gamma}^{\vartheta}_{~\xi\tau}&=&{\Gamma}^{\vartheta}_{~\xi\tau}
+\Psi^{\vartheta}_{~\xi\tau}+\Omega^{\vartheta}_{~\xi\tau},
\\\label{8a}
\Omega^{\vartheta}_{~\xi\tau}&=&\frac{1}{2}g^{\vartheta\gamma}
(Q_{\xi\tau\gamma} +Q_{\xi\tau\gamma}-Q_{\gamma\xi\tau}),
\\\label{9a}
\Psi^{\vartheta}_{~\xi\tau}&=&\tilde{\Gamma}^{\vartheta}_{~[\xi\tau]}
+g^{\vartheta\gamma}g_{\xi\varepsilon}
\tilde{\Gamma}^{\varepsilon}_{~[\tau\gamma]}+g^{\vartheta\gamma}
g_{\tau\varepsilon}\tilde{\Gamma}^{\varepsilon}_{~[\xi\gamma]},
\end{eqnarray}
where
\begin{equation}\label{10a}
Q_{\gamma\xi\tau}=\nabla_{\gamma}g_{\xi\tau}
=-g_{\xi\tau,\gamma}+g_{\tau\varepsilon}\hat{\Gamma}^{\varepsilon}_{~\xi\gamma}
+g_{\varepsilon\xi}\hat{\Gamma}^{\varepsilon}_{~\tau\gamma}.
\end{equation}
The WC framework reduces to standard Weyl geometry when the torsion
tensor vanishes as demonstrated by Eqs.(\ref{2a}) and (\ref{7a}),
where $Q_{\vartheta\xi\tau} = -2g_{\xi\tau}\Psi_{\vartheta}$.
Therefore, Eq.(\ref{7a}) becomes
\begin{equation}\label{11a}
\tilde{\Gamma}^{\vartheta}_{~\xi\tau}={\Gamma}^{\vartheta}_{~\xi\tau}
+g_{\xi\tau}\Psi^{\vartheta}
-\delta^{\vartheta}_{~\xi}\Psi_{\tau}-\delta^{\vartheta}_{~\tau}\Psi_{\xi}
+\Psi^{\vartheta}_{~\xi\tau},
\end{equation}
with
\begin{equation}\label{12a}
\Psi^{\vartheta}_{~\xi\tau}=T^{\vartheta}_{~\xi\tau}-g^{\vartheta\gamma}
g_{\varepsilon\xi}T^{\varepsilon}_{~\gamma\tau}-g^{\vartheta\gamma}
g_{\varepsilon\tau}T^{\varepsilon}_{~\gamma\xi}.
\end{equation}
The WC torsion is given by
\begin{equation}\label{13a}
T^{\vartheta}_{~\xi\tau}=\frac{1}{2}(\tilde{\Gamma}^{\vartheta}
_{~\xi\tau}-\tilde{\Gamma}^{\vartheta}_{~\tau\xi}).
\end{equation}
The WC curvature tensor can be defined in terms of connection
coefficients as
\begin{equation}\label{14a}
\tilde{S}^{\vartheta}_{~\xi\tau\gamma}=\tilde{\Gamma}^{\vartheta}
_{~\xi\gamma,\tau}-\tilde{\Gamma}^{\vartheta}_{~\xi\tau,\gamma}+\tilde{\Gamma}
^{\varepsilon}_{~\xi\gamma}
\tilde{\Gamma}^{\vartheta}_{~\varepsilon\tau}-\tilde{\Gamma}
^{\varepsilon}_{~\xi\tau}
\tilde{\Gamma}^{\vartheta}_{~\varepsilon\gamma}.
\end{equation}
The WC scalar is obtained as
\begin{eqnarray}\nonumber
\tilde{S}&=&\tilde{S}^{\xi\tau}_{~\xi\tau}
=S+6\nabla_{\tau}\Psi^{\tau}-4\nabla_{\tau}
T^{\tau}-6\Psi_{\tau}\Psi^{\tau} +8\beta
_{\tau}T^{\tau}+T^{\xi\vartheta\tau}T_{\xi\vartheta\tau}
\\\label{15a}
&+&2T^{\xi\vartheta\tau}T_{\tau\vartheta\xi}-4T^{\tau}T_{\tau}.
\end{eqnarray}

By omitting boundary terms in the Ricci scalar, the gravitational
action can be written as
\begin{equation}\label{16a}
S=\frac{1}{2\kappa} \int
g^{\xi\tau}(\Gamma^{\vartheta}_{~\gamma\xi}\Gamma^{\gamma}_{~\vartheta\tau}
-\Gamma^{\vartheta}_{~\gamma\vartheta}\Gamma^{\gamma}_{~\xi\tau})\sqrt{-g}
d^{4}x.
\end{equation}
For a symmetric connection, we have
\begin{equation}\label{17a}
\Gamma^{\vartheta}_{~\xi\tau}=-L^{\vartheta}_{~\xi\tau}.
\end{equation}
Thus, Eq.\eqref{16a} becomes
\begin{equation}\label{18a}
S=-\frac{1}{2\kappa} \int
g^{\xi\tau}(L^{\vartheta}_{~\gamma\xi}L^{\gamma}_{~\vartheta\tau}
-L^{\vartheta}_{~\gamma\vartheta}L^{\gamma}_{~\xi\tau})\sqrt{-g}
d^{4}x,
\end{equation}
where
\begin{equation}\label{19a}
Q\equiv-g^{\xi\tau}(L^{\vartheta}_{~\gamma\xi}L^{\gamma}_{~\vartheta\tau}
-L^{\vartheta}_{~\tau\vartheta}L^{\tau}_{~\xi\tau}),
\end{equation}
and
\begin{equation}\label{20a}
L^{\vartheta}_{~\xi\tau}\equiv-\frac{1}{2}g^{\vartheta\gamma}
(\nabla_{\tau}g_{\xi\gamma}+\nabla_{\xi}g_{\gamma \tau}
-\nabla_{\gamma}g_{\xi\tau}).
\end{equation}
Substituting a generic function for the non-metricity scalar in
Eq.(\ref{18a}), we have
\begin{equation}\label{21a}
S=\frac{1}{2\kappa}\int f(Q)\sqrt{-g}d^{4}x.
\end{equation}
Coupling of this action with matter-Lagrangian yields \cite{34}
\begin{equation}\label{22a}
S=\frac{1}{2\kappa}\int f(Q,L_{m}) \sqrt{-g}d^{4}x.
\end{equation}
The superpotential is given by
\begin{equation}\label{23a}
\mathcal{P}^{\vartheta}_{~\xi\tau}=-\frac{1}{2}L^{\vartheta}_{~\xi\tau}
+\frac{1}{4}(Q^{\vartheta} -\tilde{Q}^{\vartheta})g_{\xi\tau}-
\frac{1}{4} \delta ^{\vartheta} _{~[\xi Q_{\tau}]}.
\end{equation}
The non-metricity relation (provided in Appendix $\textbf{A}$) is
\begin{equation}\label{25a}
Q=-Q_{\vartheta\xi\tau}\mathcal{P} ^{\vartheta\xi\tau}=-\frac{1}{4}
(-Q^{\vartheta\xi\tau} Q_{\vartheta\xi\tau}+2Q^{\vartheta\xi\tau}
Q_{\tau\vartheta\xi} -2Q^{\vartheta}\tilde{Q}_{\vartheta}+Q
^{\vartheta}Q_{\vartheta}).
\end{equation}
The variation of Eq.(\ref{22a}) corresponding to metric tensor
yields
\begin{equation}\label{26a}
\delta S=\frac{1}{2}\int\delta [f(Q,L_{m}) \sqrt{-g}]d^{4}x
=\frac{1}{2}\int(f\delta\sqrt{-g} +(f_{Q}\delta Q+f_{L_{m}}\delta
L_{m})\sqrt{-g}d^{4}x.
\end{equation}
Moreover, we define
\begin{eqnarray}\label{27a}
\mathcal{T}_{\xi\tau}=-\frac{2}{\sqrt{-g}}\frac{\delta(\sqrt{-g}L_{m})}{\delta
g^{\xi\tau}}=g_{\xi\tau}L_{m}-2\frac{\partial L_{m}}{\partial
g^{\xi\tau}}.
\end{eqnarray}
The variation of $Q$ is given in Appendix \textbf{B}. The variation
of determinant of the metric tensor is given by
\begin{equation}\label{27aa}
\delta\sqrt{-g}=-\frac{1}{2}\sqrt{-g}g_{\xi\tau}\delta g^{\xi\tau}.
\end{equation}
Using Eqs.\eqref{27a}, \eqref{27aa} and variation of $\delta Q$ in
\eqref{26a}, it follows that
\begin{eqnarray}\nonumber
\delta S&=&\frac{-1}{2}\int f g_{\xi\tau}\sqrt{-g}\delta g^{\xi\tau}
\\\nonumber
&-&f_{Q}\sqrt{-g}(\mathcal{P}_{\xi\vartheta\gamma}
Q_{\tau}^{~\vartheta\gamma}-2Q^{\vartheta\gamma}_{~~\xi}
\mathcal{P}_{\vartheta\gamma\tau}) \delta
g^{\xi\tau}+2f_{Q}\sqrt{-g}
\mathcal{P}_{\vartheta\xi\tau}\nabla^{\vartheta} \delta g^{\xi\tau}
\\\label{28a}
&+&\frac{1}{2}f_{L_{m}}(g_{\xi\tau}L_{m}-\mathcal{T}_{\xi\tau})\sqrt{-g}
\delta g^{\xi\tau}d^ {4}x.
\end{eqnarray}
The corresponding field equations after simplification are
\begin{eqnarray}\nonumber
\frac{1}{2}f_{L_{m}}(g_{\xi\tau}L_{m}-\mathcal{T}_{\xi\tau})&=&
\frac{2}{\sqrt{-g}} \nabla_{\vartheta} (f_{Q}\sqrt{-g}
\mathcal{P}^{\vartheta}_{~\xi\tau})+ \frac{1}{2}fg_{\xi\tau}
\\\label{29a}
&+&f_{Q} (\mathcal{P}_{\xi\vartheta\gamma}
Q_{\tau}^{~\vartheta\gamma} -2Q^{\vartheta\gamma}_{~~\xi}
\mathcal{P}_{\vartheta\gamma\tau}),
\end{eqnarray}
where $f_{L_{m}}=\frac{\partial f}{\partial L_{m}}$ and
$f_{Q}=\frac{\partial f}{\partial Q}$. For
$f(Q,L_{m})=f(Q)+2{L}_{m}$, it reduces to the $f(Q)$ gravity field
equations \cite{65a}
\begin{equation}\label{29b}
-\mathcal{T}_{\xi\tau}=\frac{2}{\sqrt{-g}} \nabla_{\vartheta}
(f_{Q}\sqrt{-g} \mathcal{P}^{\vartheta}_{~\xi\tau})+
\frac{1}{2}fg_{\xi\tau}+f_{Q} (\mathcal{P}_{\xi\vartheta\gamma}
Q_{\tau}^{~\vartheta\gamma} -2Q^{\vartheta\gamma}_{~~\xi}
\mathcal{P}_{\vartheta\gamma\tau}).
\end{equation}

To investigate geometry of the CSs, we consider the interior region
as
\begin{equation}\label{8}
ds^{2}=dt^{2}e^{\nu(r)}-dr^{2}e^{\eta(r)}-r^{2}(d\theta^{2}+\sin^{2}\theta
d\phi^{2}).
\end{equation}
We assume anisotropic matter configuration with four-vector
($V_{\xi}$) and four-velocity $(U_{\xi})$ as
\begin{equation}\label{9}
\mathcal{T}_{\xi\tau}=U_{\xi}U_{\tau} \varrho + V_{\xi}V_{\tau}
p_{r}-p_{t}g_{\xi\tau} + U_{\xi}U_{\tau}p_{t} -
V_{\xi}V_{\tau}p_{t}.
\end{equation}
The resulting field equations are
\begin{eqnarray}\nonumber
\frac{1}{2}f_{L_m}(\varrho-L_m)&=&\frac{1}{2 r^{2}e^{\eta}}\bigg(2r
f_{QQ}Q'(e^{\eta}-1)+ f_{Q}((e^{\eta}-1)(2+r \nu')
\\\label{10}
&+&(1+e^{\eta})r\eta')\bigg)+\frac{f}{2},
\\\nonumber
\frac{1}{2}f_{L_m}(p_{r}-L_m)&=&\frac{-1}{2 r^{2}e^{\eta}}\bigg(2r
f_{QQ}Q'(e^{\eta}-1) +f_{Q}(e^{\eta}-1)(2+r\nu'+r\eta')
\\\label{11}
&-&2r\nu')\bigg)-\frac{f}{2},
\\\nonumber \frac{1}{2}f_{L_m}(p_{t}-L_m)&=&
\frac{-1}{4re^{\eta}}\bigg(-2rf_{QQ}Q'\nu'+
f_{Q}(2\nu'(e^{\eta}-2)-r\nu'^{2} +\eta'(2e^{\eta}
\\\label{12}
&+&r\nu')-2r\nu'')\bigg)-\frac{f}{2}.
\end{eqnarray}
These equations are complicated due to multivariate functions and
derivatives. Thus we consider specific $f(Q,L_{m})$ functional form
with arbitrary constants $\alpha$ and $\beta$ as \cite{34}
\begin{eqnarray}\label{13}
f(Q,L_{m})= 2 L_{m}-\alpha Q + \beta.
\end{eqnarray}

The chosen functional form represents a physically meaningful
extension of GR, particularly suitable for studying compact stellar
structures. In this theory, the non-metricity scalar carries the
gravitational information instead of curvature or torsion, and
taking a linear dependence on $Q$, ensuring that the field equations
remain second-order. This property avoids higher-order derivatives
that could lead to instabilities or ghost degrees of freedom, which
is crucial when analyzing static spherically symmetric systems. The
constant parameter $\alpha$ effectively modulates the strength of
gravity, reducing to GR when $\alpha=-1$ and thus allows one to
explore small deviations from GR in the strong field regime without
compromising mathematical tractability. The constant term $\beta$ in
the model serves as an effective vacuum energy or cosmological
constant-like contribution. Within dense stellar interiors, this
term may influence the pressure and density profiles, introducing a
uniform background that could mimic self-bound configurations.
Moreover, the simplicity of this linear model makes it particularly
well suited for incorporating anisotropic matter distributions in
ultra-dense stars without introducing computational complexity.
Thus, this functional form is motivated by both theoretical
consistency and astrophysical applicability.

In the context of $f(Q,L_m)$ gravity, the gravitational constant is
modified by the coupling between the matter-Lagrangian and the
torsion scalar. The parameter $\alpha$, which multiplies $Q$ plays a
critical role in modifying the gravitational interactions. In our
model, $\alpha$ directly affects the strength of the gravitational
interaction by modifying the contribution of the non-metricity term.
The parameter $\alpha$ essentially renormalizes the gravitational
strength by introducing a coupling between the matter distribution
and the non-metricity. This can be understood as a modification of
the gravitational potential, where the standard Newtonian potential
is altered by the additional term involving $\alpha$. The impact of
this renormalization is most pronounced in stellar environments
where the matter density is high and the effects of modified gravity
are more significant.

The constant term like $\beta$, which has the characteristics of a
cosmological constant, is associated with the large-scale behavior
of spacetime. However, in the context of compact star interiors,
where spacetime is highly curved and the matter distribution is
dense, the influence of such a constant background term is not as
straightforward. Specifically, the contribution of $\beta$ to the
metric inside a compact star could be diluted by the dynamic
interactions between the gravitational field and the matter
distribution, making it less relevant in the strongly curved
regions. To better understand the role of $\beta$ in the interior of
a compact star, the term $\beta$ modifies the spacetime curvature
inside the compact star. Since the interior is highly curved due to
the dense matter, we expect that the influence of a constant term
like $\beta$ could be less significant compared to the local matter
contributions. However, the exact contribution of $\beta$ will
depend on the specific stellar configuration and we will investigate
this through detailed metric calculations.

The linear model is the minimal and physically transparent extension
of symmetric teleparallel gravity that includes matter-geometry
coupling. The coupling term $\alpha$ controls how strongly
non-metricity affects gravitational dynamics, directly impacting
mass, radius, redshift and stability. While, the term $\beta$ acts
as a background shift, influencing compactness, pressure balance and
allowable stellar configurations. Both these parameters are
constrained using astrophysical data from compact stars,
gravitational waves and stability analysis. This model represents
one of the simplest yet physically meaningful extensions of
symmetric teleparallel gravity. This linear choice allows us to
retain analytical tractability while capturing deviations from GR
through the coupling constant $\alpha$ and the cosmological-like
term $\beta$. We choose the value of matter-Lagrangian density as
$L_{m}=p_r$ \cite{37k}. The field equations corresponding to this
model become
\begin{eqnarray} \nonumber
\varrho&=&\frac{1}{2} \bigg[ \beta -2(\beta+\frac{e^{-\eta}
(-\alpha(e^{\eta}-1)(r (\eta'+\nu')+2)-2 r \nu')}{r^2}
\\\nonumber
&-&\frac{\alpha(e^{-\eta}-1)(\eta'+\nu')}{r})+\frac{\alpha e^{-\eta}
(r(e^{\eta}+1) \eta'+(e^{\eta}-1)(r \nu'+2))}{r^2}
\\\label{13aa}
&+&\frac{\alpha(e^{-\eta}-1)(\eta'+\nu')}{r}\bigg],
\\\nonumber
p_{r}&=&\frac{1}{2} \bigg[-\beta -\frac{e^{-\eta}(-\alpha(e^{\eta
(r)}-1)(r (\eta'+\nu')+2)-2 r \nu')}{r^2}
\\\label{13bb}
&+&\frac{\alpha(e^{-\eta}-1)(\eta'+\nu')}{r}\bigg],
\\\nonumber
p_{t}&=&\frac{1}{4 r}\bigg[e^{-\eta} (2
e^{\eta}(\alpha(e^{-\eta}-1)(\eta'+\nu')+\alpha (\eta'+\nu')-\beta
r))
\\\label{13cc}
&-& 2 \alpha r \nu'' +\alpha  \nu'(r (\eta'- \nu')-4)\bigg].
\end{eqnarray}

\section{Analysis of Viable Solutions}

Here, we investigate two well-motivated non-singular solutions that
are effective in astrophysical system. Detailed mathematical
derivations and physical interpretations of both solutions are
provided in the given sections. By examining these solutions, we
determine whether cosmic objects will maintain structural integrity
over time. By integrating the non-singular solutions into our
investigation, we seek to advance our comprehension of the physical
mechanisms that demonstrate the celestial structure and evolution.

\subsection{Solution I: The Finch-Skea Spacetime}

The Finch-Skea solution is a nonsingular and analytic stellar model
that balances physical realism with mathematical simplicity, making
it a versatile tool for probing compact objects in both standard and
modified gravity. Its design elegantly balances mathematical
tractability with key physical requirements of regularity at the
origin, a monotonically decreasing density profile, finite central
pressure and density as well as seamless matching to the
Schwarzschild exterior. By employing a simple polynomial function
for the metric, it reduces the complexity of the Einstein field
equations to closed form expressions for all thermodynamic and
geometric variables. This analytic transparency allows for direct
scrutiny of stability criteria, energy conditions and causality,
making it an invaluable pedagogical and benchmarking tool.
Furthermore, its adaptable framework readily extends to anisotropic
pressures and modified theories of gravity, where it serves as a
foundational ansatz to isolate and interpret deviations from general
relativity. In an era dominated by numerical simulations, the
Finch-Skea model remains a cornerstone for theoretical insight and a
practical bridge linking geometric assumptions to compact star
properties. We consider viable non-singular Finch-Skea solution with
arbitrary constants $(a,s,c)$ as \cite{37a}
\begin{eqnarray}\label{14}
e^{\nu}=(a+\frac{s}{2}r\sqrt{c r^{2}})^{2},  \quad e^{\eta}=1+c
r^{2}.
\end{eqnarray}
The Finch-Skea solution provides a robust mathematical foundation
for analyzing the behavior of celestial objects under extreme
conditions. The constants are determined by Darmois junction
conditions. The corresponding field equations are
\begin{eqnarray}\nonumber
\varrho&=&\frac{-1}{2(1+r^2z)^2(r^3sc+2a\sqrt{r^2c})}\bigg[r^7sc^3
\beta-16rsc+2a(r^2c)^{\frac{5}{2}}\beta\\\nonumber &+&r^3sc(2c
(\beta-8 \alpha)+2r^5 s c^2(\beta-c \alpha)+4 a (r^2
c)^{\frac{3}{2}}(\beta-c \alpha)\\\label{16} & +& 2 a \sqrt{r^2c})
(2 c \alpha + \beta) \bigg],
\\\nonumber
p_{r}&=&\frac{1}{2 (1 + r^2 c) (r^3 s c + 2 a \sqrt{r^2c})}\bigg[8 r
s c + r^3 s c (2 c \alpha - \beta) + 2 a \sqrt{r^2c} (2 c \alpha
\\\label{17}
&-& \beta) - r^5 s c^2 \beta-2 a (r^2 c)^{\frac{3}{2}} \beta \bigg],
\\\nonumber
p_{t}&=&\frac{-1}{2 (1 + r^2 c) (r^3 s c + 2 a \sqrt{r^2c})}\bigg[8
r s c \alpha + 2 r^5 s c^2 \beta + r^7 s c^3 \beta + 4 a (r^2
c)^{\frac{3}{2}} \beta
\\\label{18}
&+& 2 a (r^2 c)^{\frac{5}{2}} \beta + 2 a\sqrt{r^2c} (-2 c \alpha +
\beta) + r^3 s c (2 c \alpha + \beta)\bigg].
\end{eqnarray}

The Schwarzschild solution describes the spacetime geometry
generated by a massive, non-rotating, spherically symmetric object
and is one of the simplest and most useful exact solutions of the
Einstein field equations. In its derivation, the spacetime is
assumed to be vacuum, spherically symmetric, and static. However,
the assumption of static is in fact redundant by Birkhoff theorem,
where any spherically symmetric vacuum solution of the Einstein
field equations must be stationary. As a consequence, the exterior
spacetime of any spherically symmetric compact star remains static,
implying that such an object does not emit gravitational waves as
long as spherical symmetry is preserved. Imposing spherical
symmetry, static and the vacuum condition significantly restricts
the general form. A spherically symmetric spacetime is invariant
under spatial rotations and reflections. A static spacetime is
characterized by metric components that are independent of the time
coordinate and by invariance under time reversal. We assume the
external geometry of the stellar interiors as
\begin{eqnarray}\label{15}
ds^{2}_{+}=dt^{2}\aleph-d
r^{2}\aleph^{-1}-r^{2}(d\theta^{2}+\sin^{2}\theta d\phi^{2}),
\end{eqnarray}
with $\aleph=\left(1-\frac{2M}{r}\right)$. A single solution to the
field equations does not adequately define the geometry of an object
in relativistic scenarios.

In general, the interior of a stellar configuration is represented
by a static spherically symmetric spacetime and the exterior is
characterized by the Schwarzschild vacuum solution. To ensure smooth
transition between these two areas, specific limitations known as
matching conditions must be applied. Consequently, adhering to these
conditions guarantees that the physical characteristics remain
consistent across the hypersurface, thereby creating a unified
description of spacetime. The selection of the external solution
should align with the behavior of the interior geometry whether it
is static or dynamic and regardless of charge to maintain
consistency at the matching interface. For a metric function to
satisfy the Darmois junction conditions, it must be continuously
differentiable. By enforcing this condition, a smooth solution can
be achieved by equating the metric potentials of the interior and
exterior at boundary $\Sigma$ (where $r=\mathbf{R}$) as
\begin{equation}\label{3.3.4}
\mathfrak{g}_{tt}^-=\mathfrak{g}_{tt}^+,\quad
\mathfrak{g}_{rr}^-=\mathfrak{g}_{rr}^+,\quad
\mathfrak{g}_{tt,r}^-=\mathfrak{g}_{tt,r}^+.
\end{equation}
At the surface boundary $(r=\mathbf{R})$, we get
\begin{eqnarray}\label{16}
\mathfrak{g}_{tt}&=& [a+\frac{s}{2}\mathbf{R}\sqrt{c
\mathbf{R}^{2}}]^{2}=1-\frac{2M}{\mathbf{R}},
\\\label{17}
\mathfrak{g}_{rr}&=&1+c
\mathbf{R}^{2}=(1-\frac{2M}{\mathbf{R}})^{-1},
\\\label{18}
\mathfrak{g}_{tt,r}&=& s\mathbf{R}\sqrt{c}(a+\frac{s}{2}
\mathbf{R}\sqrt{c \mathbf{R}^{2}})=\frac{M}{\mathbf{R}^{2}}.
\\\nonumber
\end{eqnarray}
Solving the above equations simultaneously, we have
\begin{eqnarray}\label{18aa}
a=\frac{2\mathbf{R}-5M}{2\sqrt{\mathbf{R}^{2}-2M\mathbf{R}}}, \quad
s=\frac{1}{\mathbf{R}}\sqrt{\frac{M}{2\mathbf{R}}}, \quad
c=\frac{2M}{\mathbf{R}^{2}(\mathbf{R}-2M)}.
\end{eqnarray}
For a singularity-free spacetime, the metric functions must be
regular. The calculated values of stellar mass and radius are shown
in Table \textbf{1} \cite{38}-\cite{44}. The values of constants
corresponding to mass and radius of the stellar interior for
solution \textbf{I} are demonstrated in Table \textbf{2}. In Figure
\textbf{1}, the metric coefficients of Finch-Skea solution exhibit
the necessary regularity and increasing behavior.
\begin{table}\caption{Estimation of mass and radius of stellar
interior.}
\begin{center}
\begin{tabular}{|c|c|c|}
\hline Stellar Interiors & $M(M_{\odot})$ & $\mathbf{R}(km)$
\\
\hline 4U 1538-52 & 0.87 $\pm$ 0.07 & 7.866 $\pm$ 0.21
\\
\hline Her X-1 & 0.85 $\pm$ 0.15 & 8.1 $\pm$ 0.41
\\
\hline LMC X-4 & 1.04 $\pm$ 0.09 & 8.301 $\pm$ 0.2
\\
\hline 4U 1820-30 & 1.58 $\pm$ 0.06 & 9.1 $\pm$ 0.4
\\
\hline Cen X-3 & 1.49 $\pm$ 0.08 & 9.178 $\pm$ 0.13
\\
\hline 4U 1608-52 & 1.74 $\pm$ 0.01 & 9.3 $\pm$ 0.10
\\
\hline PSR J1903+327 & 1.667 $\pm$ 0.021 & 9.48 $\pm$ 0.03
\\
\hline PSR J1614-2230 & 1.97 $\pm$ 0.04 & 9.69 $\pm$ 0.2
\\
\hline Vela X-1 & 1.77 $\pm$ 0.08 & 9.56 $\pm$ 0.08
\\
\hline EXO 1785-248 & 1.30 $\pm$ 0.2 & 10.10 $\pm$ 0.44
\\
\hline SMC X-4 & 1.29 $\pm$ 0.05 & 8.831 $\pm$ 0.09
\\
\hline
\end{tabular}
\end{center}
\end{table}
\begin{table}\caption{Estimation of constant parameters for solution \textbf{I}.}
\begin{center}
\begin{tabular}{|c|c|c|c|}
\hline Stellar Interiors & a & s $(km)^{-1}$ & c $(km)^{-2}$
\\
\hline 4U 1538-52 & 0.721647 & 0.0362963 & 0.00781918
\\
\hline Her X-1 & 0.737986 & 0.0343333 & 0.00682713
\\
\hline LMC X-4 & 0.677867 & 0.0366062 & 0.00849917
\\
\hline 4U 1820-30 & 0.515527 & 0.0393097 & 0.0126622
\\
\hline Cen X-3 & 0.556388 & 0.0376881 & 0.0108966
\\
\hline 4U 1608-52 & 0.463743 & 0.0399286 & 0.0142208
\\
\hline PSR J1903+327 & 0.507239 & 0.0379742 & 0.0119768
\\
\hline PSR J1614-2230 & 0.396269 & 0.0399467 & 0.0159309
\\
\hline Vela X-1 & 0.471463 & 0.0386397 & 0.013149
\\
\hline EXO 1785-248 & 0.667333 & 0.0304946 & 0.00599414
\\
\hline SMC X-4 & 0.611888 & 0.0371547 & 0.00969829\\
\hline
\end{tabular}
\end{center}
\end{table}
\begin{figure}
\epsfig{file=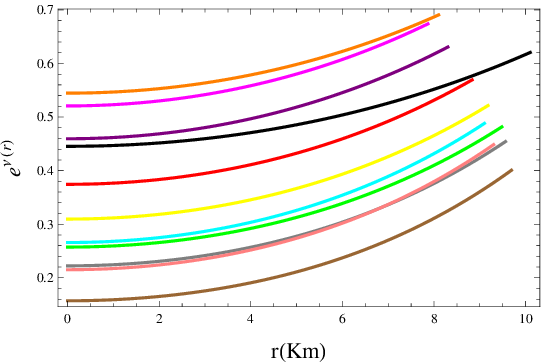,width=.45\linewidth}
\epsfig{file=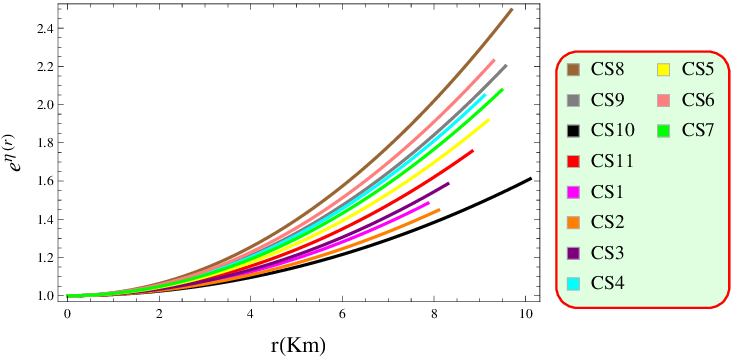,width=.6\linewidth}\caption{Plots of metric
potentials against radial coordinate for solution \textbf{I}.}
\end{figure}

\subsection{Solution II: Karmarkar Condition}

The Karmarkar condition, formulated by Shanti Swarup Karmarkar in
the 1940s, is a geometric embedding constraint expressed through the
curvature tensor. It serves as a powerful mathematical tool for
understanding spacetime structure, with applications ranging from
deriving exact solutions and analyzing singularities to modeling
cosmological scenarios. In relativistic astrophysics, imposing this
condition on static spherically symmetric spacetimes reduces the
degrees of freedom in the metric, enabling the generation of
tractable analytic models for compact stars. This approach
facilitates the study of internal pressure-density profiles and
stability against gravitational collapse of CSs. By providing a
geometric closure relation, the Karmarkar ansatz allows the
construction of physically admissible stellar solutions even in
non-Riemann frameworks, making it a valuable technique for exploring
how alternative gravity theories affect their structure and
stability.

To study cosmological models and specific properties of spacetime, a
well-known constraint is used named as Karmarkar condition
\cite{44a}. This condition incorporates curvature tensor to analyze
spacetime curvature. A key advantage of employing this condition is
its capacity to elucidate the relationship between pressure-density
distributions and overall stability of celestial objects. Through
its application, we can refine theoretical predictions and enhance
the interpretation of observational data, thereby improving our
understanding of celestial dynamics. Thus, this condition represents
a significant tool for assessing the stability of astrophysical
structures, offering critical insights into their evolution and
dynamical behavior. The Karmarkar condition is defined as
\begin{eqnarray}\label{39a}
\mathbf{R}_{1414}\mathbf{R}_{2323}=\mathbf{R}_{1212}\mathbf{R}_{3434}
+\mathbf{R}_{1224}\mathbf{R}_{1334},
\end{eqnarray}
where
\begin{eqnarray}\nonumber
\mathbf{R}_{1414}&=&-\frac{1}{4}(2\nu''+\nu'^{2}-\nu'\eta')e^{\nu},
\quad
\mathbf{R}_{2323}=\frac{1}{e^{\eta}}r^{2}\sin^{2}\theta(e^{\eta}-1),
\\\nonumber
\mathbf{R}_{3434}&=&\frac{-1}{2}r^{2}\sin^{2}\theta
\nu'e^{\nu-\eta}, \quad \mathbf{R}_{1212}=\frac{1}{2}r\eta', \quad
\mathbf{R}_{1334}=\mathbf{R}_{1224}\sin^{2}\theta=0.
\end{eqnarray}
Substituting these values in Eq.(\ref{39a}), we have
\begin{equation}\label{40a}
\nu'(\eta'-\nu')-2\nu''=\frac{\nu'\eta'}{1-e^{\eta}},
\end{equation}
where $e^{\eta}\neq 1$. Integrating the above equation, we have
\begin{equation}\label{41a}
e^{\nu(r)}=(\imath+\sigma \int\sqrt{e^{\eta}-1}dr)^{2}.
\end{equation}
We attempted to choose the metric element $e^{\eta}$ with value $-4$
\cite{37l} to obtain viable results in the $f({Q},{L}_{m})$ theory
as
\begin{equation}\label{42a}
e^{\eta(r)}=1+(1 + B r^{2})^{-4}A^{2} r^{2}.
\end{equation}
Manipulate Eqs.(\ref{41a}) and (\ref{42a}), we obtain
\begin{eqnarray}\label{43a}
e^{\nu(r)}=\bigg(\imath-\frac{A \sigma}{2B(1+B r^{2})}\bigg)^{2},
\end{eqnarray}
where $(A,B,\sigma,\imath)$ are the arbitrary constants. The field
equations for the Karmarkar solution are
\begin{eqnarray}\nonumber
\varrho&=&\frac{1}{2 (2 \imath B (1 + B r^2)-A \sigma) (A^2 r^2 + (1
+ B r^2)^4)^2}\bigg[-2 \imath B (1 + B r^2)^9 \beta
\\\nonumber
&+& A^5 \sigma r^2 (-2 \alpha + r^2 \beta) - 2 A^4 \imath B r^2 (1 +
B r^2) (-2 \alpha + r^2 \beta)- 4 A^2 \imath B (1
\\\nonumber
&+& B r^2)^4 (\alpha - 7 B r^2 \alpha + r^2 \beta + B r^4 \beta) + A
\sigma (1 + B r^2)^7 (\beta + B (16
\\\label{49a}
&+& r^2 \beta)) +   2 \sigma (A + A B r^2)^3 (\alpha + r^2 (\beta +
B (8 - 7 \alpha + r^2 \beta)))\bigg],
\\\nonumber
p_{r}&=&\frac{1}{2 (2 \imath B (1 + B r^2)-A \sigma) (A^2 r^2 + (1 +
B r^2)^4)^2}\bigg[2 A (4 B \sigma (1 + B r^2)^3
\\\nonumber
&+& A (-A \sigma + 2 \imath B (1 + B r^2)) \alpha) + (A \sigma - 2
\imath B (1 + B r^2)) (A^2 r^2 + (1
\\\label{50a}
&+& B r^2)^4) \beta\bigg],
\\\nonumber
p_{t}&=& \frac{1}{2 (2 \imath B (1 + B r^2)-A \sigma) (A^2 r^2 + (1
+ B r^2)^4)^2}\bigg[A^5 \sigma r^4 \beta -   2 A^4 \imath B r^4 (1
\\\nonumber
&+& B r^2) \beta - 2 \imath B (1 + B r^2)^9 \beta +   2 \sigma (A +
A B r^2)^3 (-\alpha + B r^2 \alpha + r^2 \beta
\\\nonumber
&+& B r^4 \beta) -  4 A^2 \imath B (1 + B r^2)^4 (-\alpha + 3 B r^2
\alpha + r^2 \beta + B r^4 \beta) + A \sigma (1
\\\label{51a}
&+& B r^2)^6 (8 B (-1 + B r^2) \alpha + (1 + B r^2)^2 \beta)\bigg].
\end{eqnarray}
The unknown constants are calculated by the first Darmois junction
condition as
\begin{eqnarray}\label{48a}
A&=&\frac{(1+B \mathbf{R}^2)^2}{\mathbf{R}}
\big(\frac{2M}{\mathbf{R}-2M}\big)^{1/2},
\\\label{50}
B&=&\frac{(4M-\mathbf{R})}{\mathbf{R}^2(9\mathbf{R}-20M)},
\\\label{51}
\imath&=&\frac{A
\sigma}{2B(1+B\mathbf{R}^2)}+\big(1-\frac{2M}{\mathbf{R}}\big)^{1/2},
\\\label{52}
\sigma&=&\frac{1}{2\mathbf{R}}\big(\frac{2M}{\mathbf{R}}\big)^{1/2}.
\end{eqnarray}
For a singularity free spacetime, the metric functions must be
regular. In Figure \textbf{2}, the metric coefficients of Karmarkar
solution exhibit the necessary regularity and increasing behavior
for solution \textbf{II}. The values of constants corresponding to
mass and radius of stellar interior for solution \textbf{II} are
represented in Table \textbf{3}.
\begin{figure}
\epsfig{file=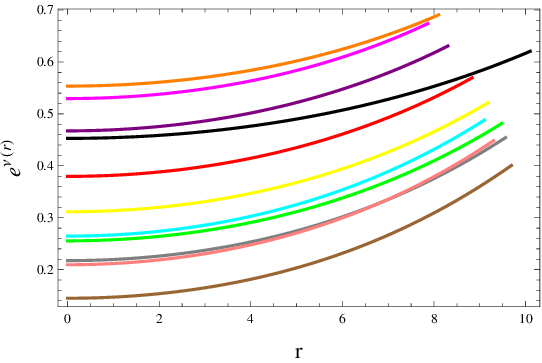,width=.45\linewidth}
\epsfig{file=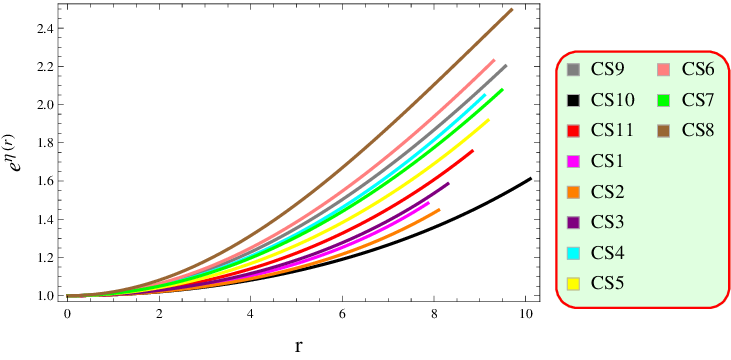,width=.6\linewidth}\caption{Plots of metric
elements against radial coordinate for solution \textbf{II}.}
\end{figure}
\begin{table}\caption{Estimation of constant parameters for solution \textbf{II}.}
\begin{center}
\begin{tabular}{|c|c|c|c|c|}
\hline Stellar Interiors & $ A(km)^{-1}$ & $B (km)^{-2}$ & $\imath$
& $\sigma (km)^{-1}$
\\
\hline 4U 1538-52  & 0.0665451 & -0.000454051 & -2.02501 &0.0304946
\\
\hline Her X-1  & 0.0728604 & -0.000983907 & -0.606793 & 0.0343333
\\
\hline LMC X-4 & 0.080114 & -0.000714639 & -1.22999 & 0.0366062
\\
\hline 4U 1820-30 & 0.113904 & 0.0000737372 & 0.0393097 & 0.0393097
\\
\hline Cen X-3 & 0.105688 & -0.000120615 & -16.6771 & 0.0376881
\\
\hline 4U 1608-52 & 0.116776 & 0.000342185 & 6.91563 & 0.0399286
\\
\hline PSR J1903+327 & 0.116273 & 0.000107231 & 22.1351 & 0.0379742
\\
\hline PSR J1614-2230 & 0.0839689 & -0.000929059 & -0.647774 &
0.0363264
\\
\hline Vela X-1 & 0.129893 & 0.000283046 & 9.60884 & 0.0386397
\\
\hline EXO 1785-248 & 0.0665451 & -0.000454051 & -2.02501& 0.102876
\\
\hline SMC X-4 & 0.102876 & -0.00037901 & -4.64919 & 0.0371547
\\
\hline
\end{tabular}
\end{center}
\end{table}

\section{Feasible Characteristics of Compact Stars}

A viable stellar model must meet the essential physical and
mathematical criteria to be considered as realistic. These criteria
guarantee that the solution is meaningful in a physical sense,
aligning with a stable and consistent structure. An acceptable model
needs to ensure that the metric potentials should be free of
singularities, providing a smooth and clearly defined geometry.
Additionally, thermodynamic quantities, like energy density and
pressure remain positive and finite throughout the star's interior
and adhere the energy conditions (ECs). Moreover, stability analysis
is significant to determine that the configuration can hold its
equilibrium despite small disturbances. Thus, these factors
establish the physical validity of a stellar model which is
essential for its relevance in astrophysics.

The model parameters characterize the deviation of the $f(Q, L_m)$
model from GR and play a central role in shaping the physical
properties of the stellar interior. In this analysis, we select
small perturbations around unity (e.g., $\alpha=-0.5,-0.7$) to
ensure a physically meaningful deviation that preserves the overall
GR structure while allowing measurable corrections. Such a range
maintains the weak-field correspondence and ensures the field
equations remain well-posed and stable. The chosen parameter values
were constrained by the regularity of matter variables, as well as
by the satisfaction of the standard energy conditions. Only
combinations of $\alpha=-0.5,~\beta=-0.2$ and
$\alpha=-0.7,~\beta=-0.1$ yield finite density, positive pressures.
Values that produced unphysical negative pressures or divergent
curvature invariants were excluded. Thus, the considered parameter
values were physically motivated and systematically constrained
through theoretical limits ensuring reduction to GR, regularity and
energy condition checks and observational compatibility with
realistic compact objects. The small deviations from unity provide
stable and well-behaved stellar configurations, making them the most
suitable numerical choices for graphical representation.

\subsection{Analysis of Matter Contents}

The physical properties of matter contents like density, radial and
tangential pressures play a fundamental role in self-gravitating
systems \cite{61a}. The interplay between these components
demonstrates the crucial balance between gravity and pressure in
CSs. Figures \textbf{3}-\textbf{6} show the behavior of density,
pressure components and their rates of change for each CS. These
graphs show that the matter contents are positive and maximum at the
center. Furthermore, the rate of change of matter contents are zero
and negative at the center. Both the density and pressure components
and also their rates of change satisfy the feasibility of dense CSs
for both solutions.
\begin{figure}\center
\epsfig{file=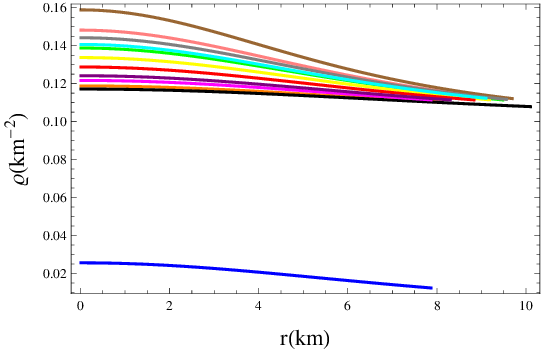,width=.45\linewidth}\epsfig{file=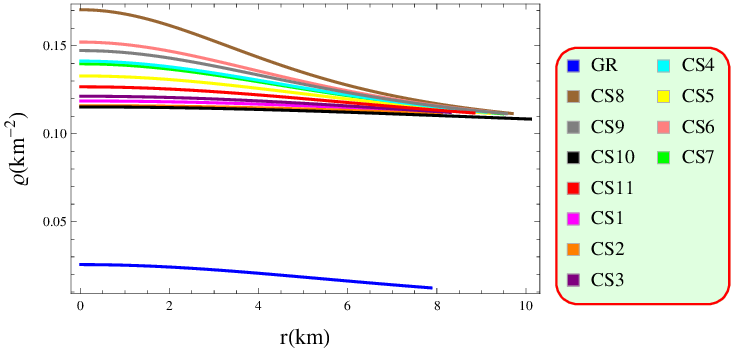,width=.6\linewidth}
\epsfig{file=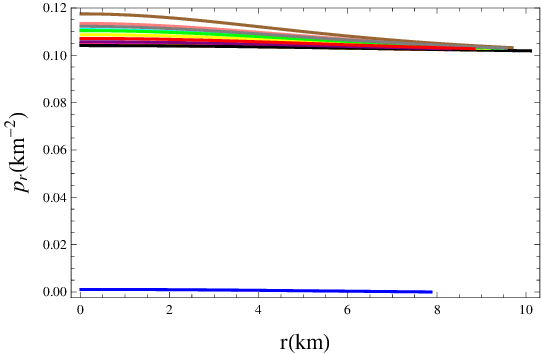,width=.45\linewidth}\epsfig{file=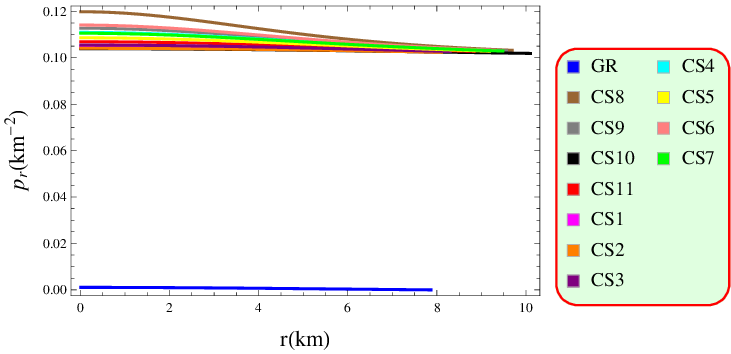,width=.6\linewidth}
\epsfig{file=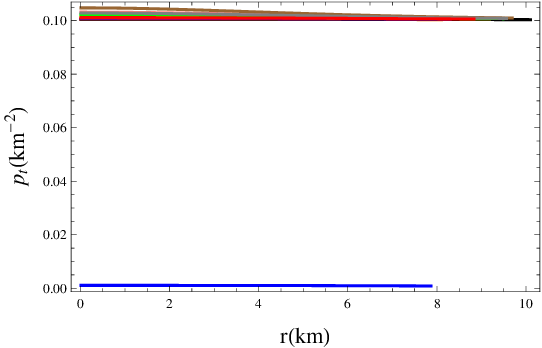,width=.45\linewidth}\epsfig{file=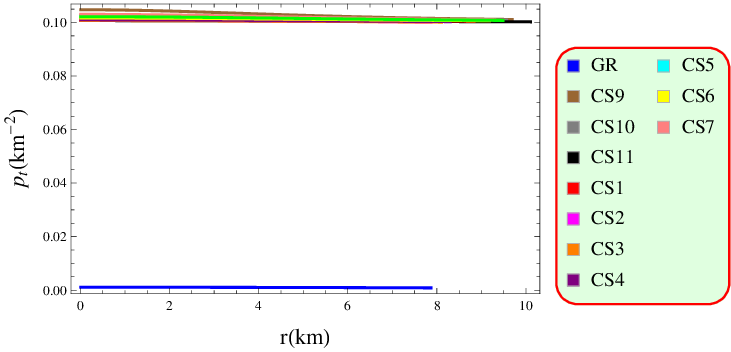,width=.6\linewidth}
\caption{Matter contents against radial coordinate for solutions
\textbf{I}(left) and \textbf{II}(right) for $\alpha=-0.5$ and
$\beta=-0.2$.}
\end{figure}
\begin{figure}\center
\epsfig{file=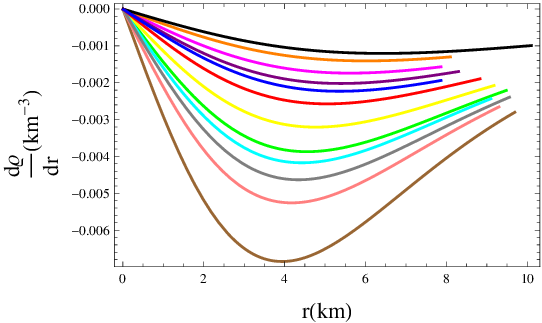,width=.45\linewidth}\epsfig{file=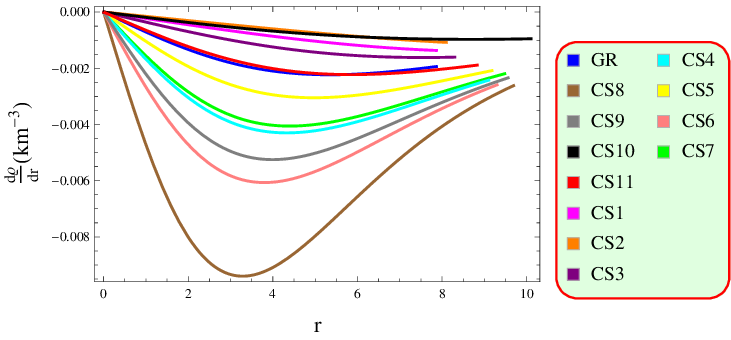,width=.6\linewidth}
\epsfig{file=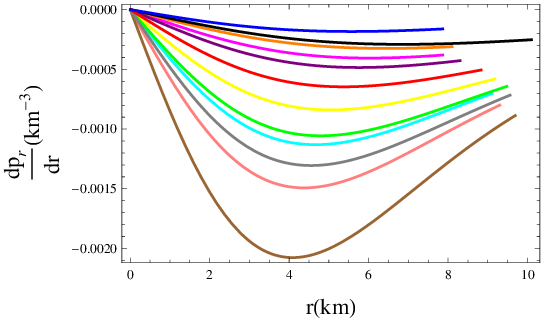,width=.45\linewidth}\epsfig{file=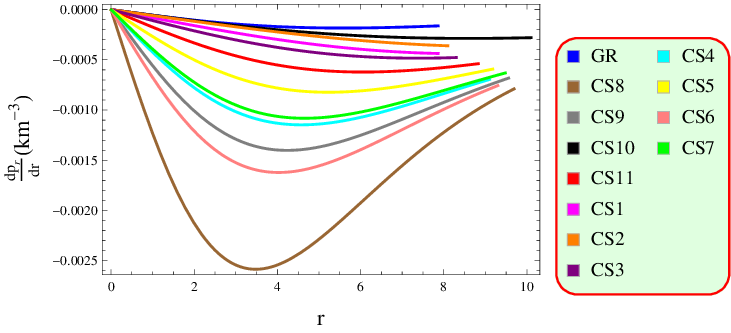,width=.6\linewidth}
\epsfig{file=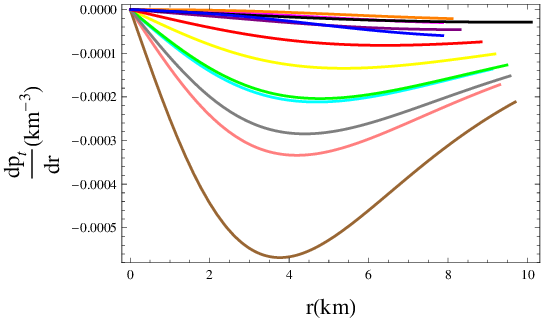,width=.45\linewidth}\epsfig{file=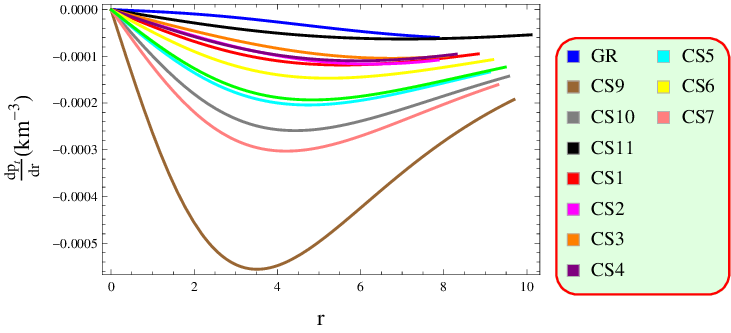,width=.6\linewidth}
\caption{Matter content's gradient against radial coordinate for
solutions \textbf{I}(left) and \textbf{II}(right) for $\alpha=-0.5$
and $\beta=-0.2$.}
\end{figure}
\begin{figure}\center
\epsfig{file=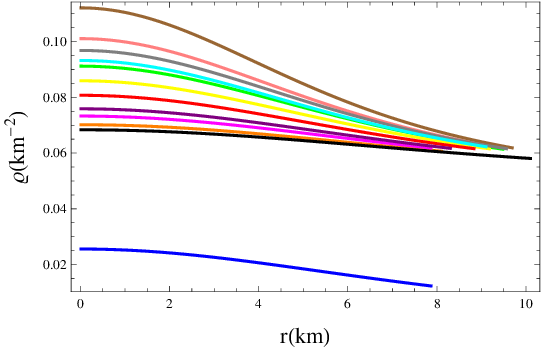,width=.45\linewidth}\epsfig{file=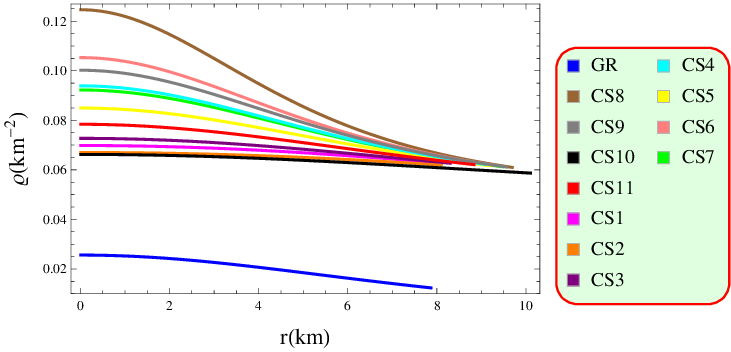,width=.6\linewidth}
\epsfig{file=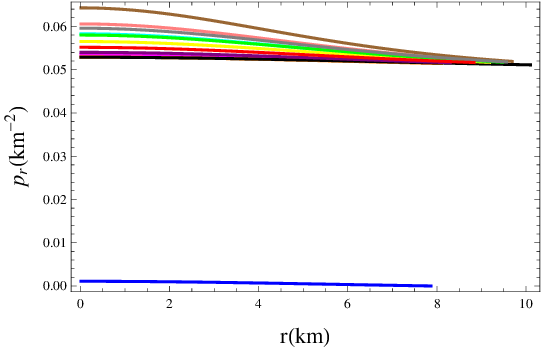,width=.45\linewidth}\epsfig{file=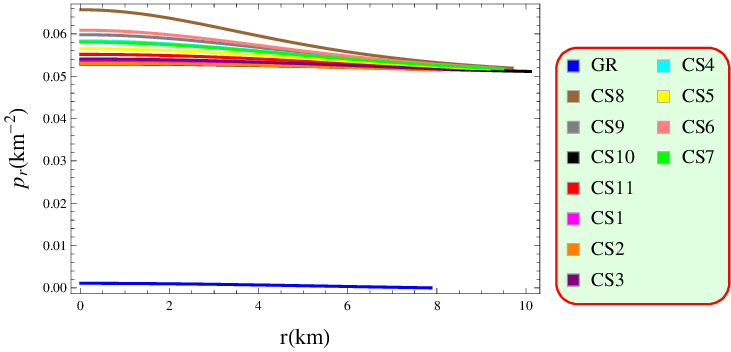,width=.6\linewidth}
\epsfig{file=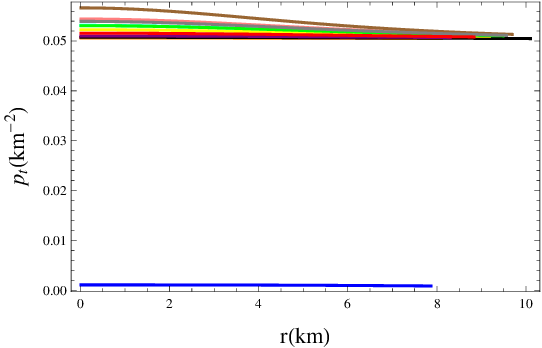,width=.45\linewidth}\epsfig{file=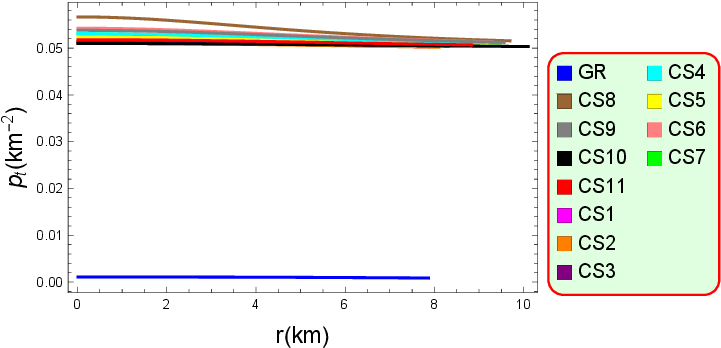,width=.6\linewidth}
\caption{Matter contents against radial coordinate for solutions
\textbf{I}(left) and \textbf{II}(right) for $\alpha=-0.7$ and
$\beta=-0.1$.}
\end{figure}
\begin{figure}\center
\epsfig{file=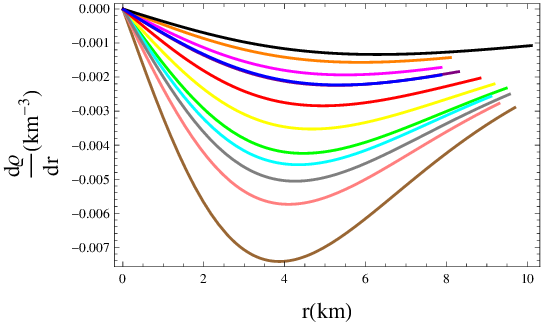,width=.45\linewidth}\epsfig{file=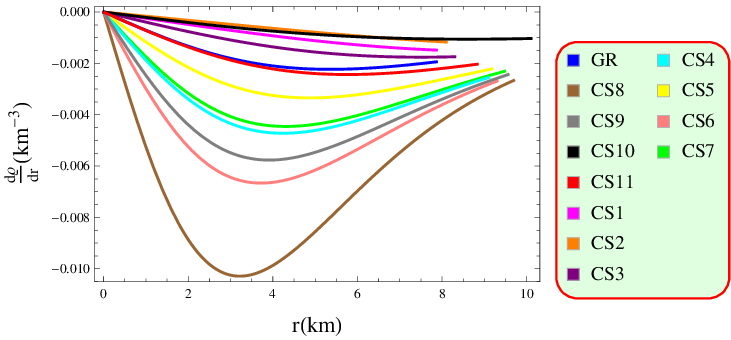,width=.6\linewidth}
\epsfig{file=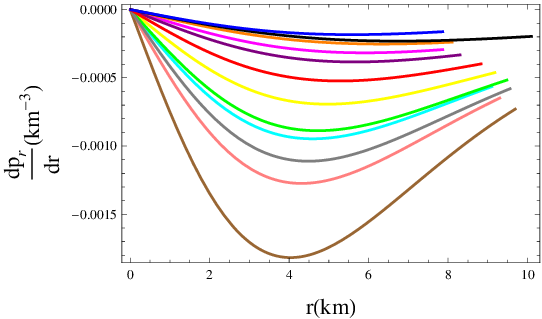,width=.45\linewidth}\epsfig{file=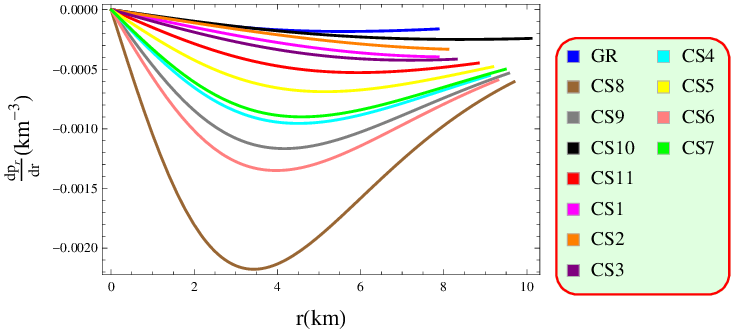,width=.6\linewidth}
\epsfig{file=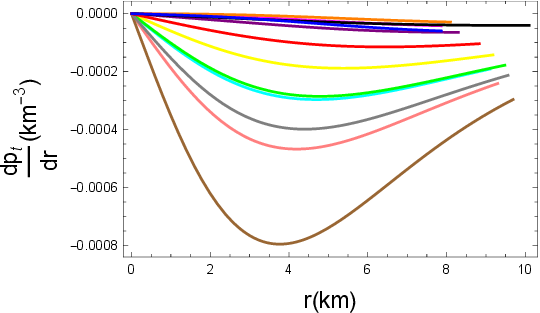,width=.45\linewidth}\epsfig{file=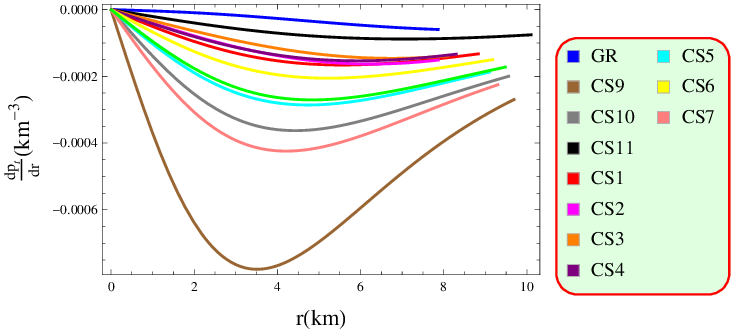,width=.6\linewidth}
\caption{Matter content's gradient against radial coordinate for
solutions \textbf{I}(left) and \textbf{II}(right) for $\alpha=-0.7$
and $\beta=-0.1$.}
\end{figure}

\subsection{Behavior of Energy Constrains}

Astrophysical objects consist of various materials in their
configuration, making it essential to distinguish between different
types of substances (ordinary or exotic) in CSs. Energy conditions
are crucial to understand how matter and energy respond to
gravitational influence. These constraints relate energy density
with pressure in spacetime and describe field properties that align
with gravitational theory. These energy constraints are defined into
four categories as dominant EC $(0\leq \varrho\pm p_{r},~ 0\leq
\varrho\pm p_{t})$, null EC $(0\leq p_{r}+\varrho, ~0\leq
p_{t}+\varrho)$, weak EC $(0\leq p_{r}+\varrho,~ 0\leq
p_{t}+\varrho, ~0\leq \varrho)$ and strong EC $(0\leq
p_{r}+\varrho,~ 0\leq p_{t}+\varrho, ~ 0\leq p_{r}+2p_{t}+\varrho)$.
By investigating ECs, we uncover insights into cosmic configurations
and how they relate to the stress-energy tensor. Since stellar
structures consist of ordinary matter, it is essential that the
parameters characterizing the interiors of CSs adhere to these ECs.
Figures \textbf{7} and \textbf{8} demonstrate the positive behavior
of ECs for the CSs. The positive energy conditions for anisotropic
CSs indicate presence of ordinary matter.
\begin{figure}\center
\epsfig{file=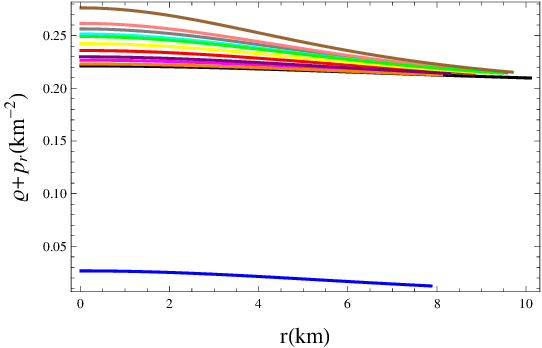,width=.45\linewidth}\epsfig{file=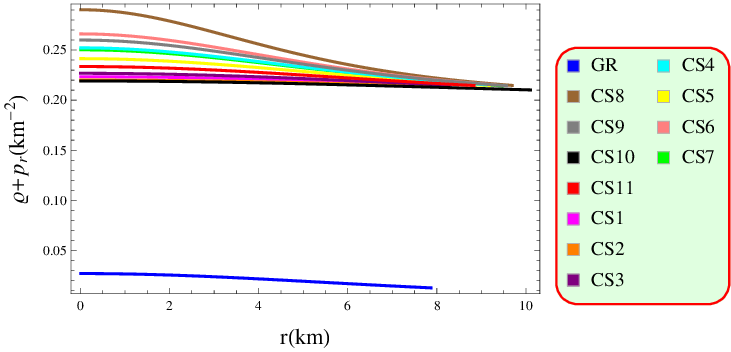,width=.6\linewidth}
\epsfig{file=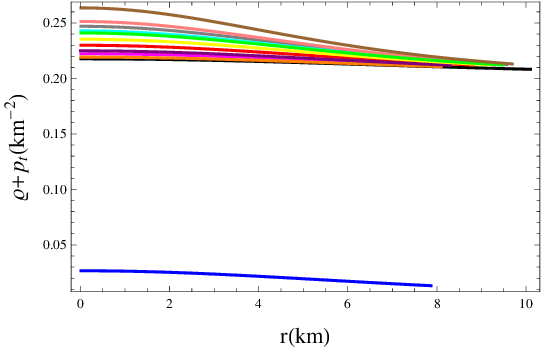,width=.45\linewidth}\epsfig{file=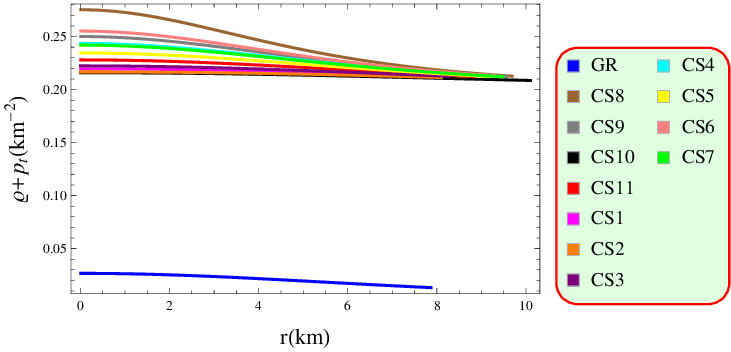,width=.6\linewidth}
\epsfig{file=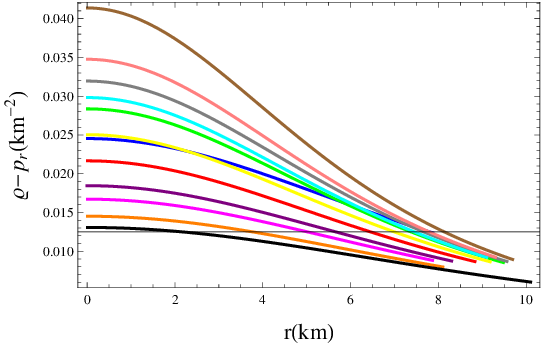,width=.45\linewidth}\epsfig{file=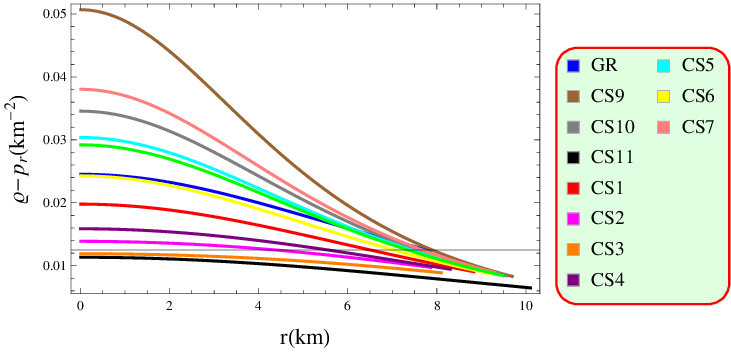,width=.6\linewidth}
\epsfig{file=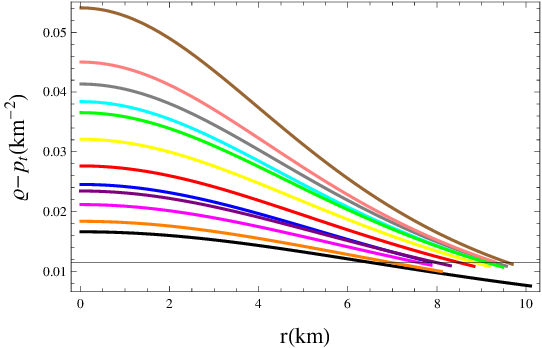,width=.45\linewidth}\epsfig{file=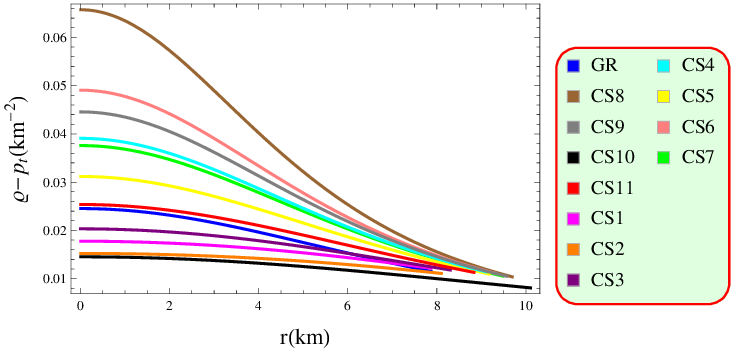,width=.6\linewidth}
\caption{Energy bounds against radial coordinate for solutions
\textbf{I}(left) and \textbf{II}(right) for $\alpha=-0.5$ and
$\beta=-0.2$.}
\end{figure}
\begin{figure}\center
\epsfig{file=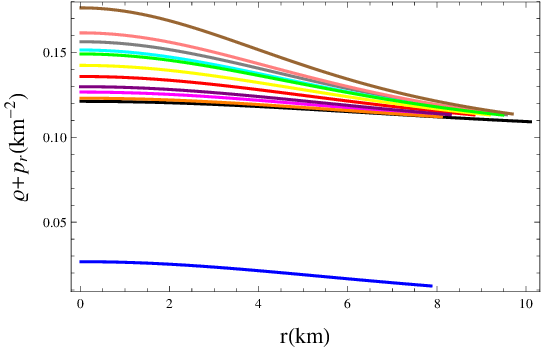,width=.45\linewidth}\epsfig{file=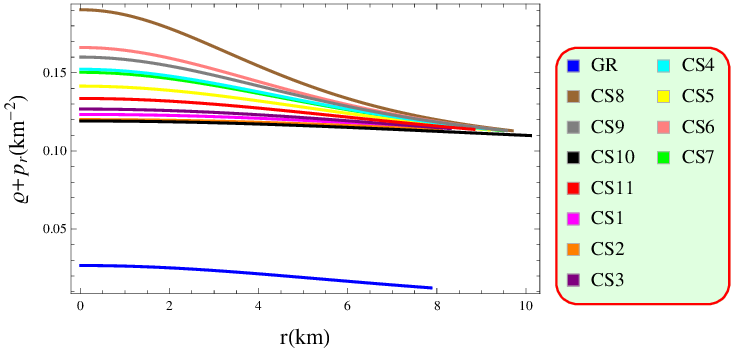,width=.6\linewidth}
\epsfig{file=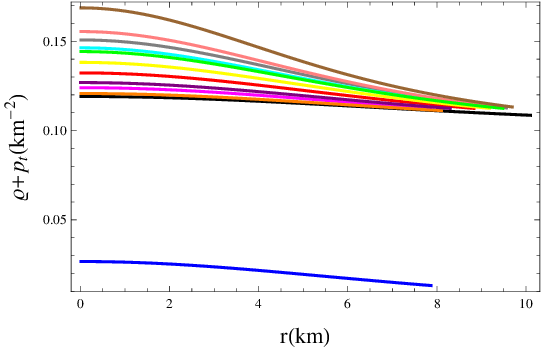,width=.45\linewidth}\epsfig{file=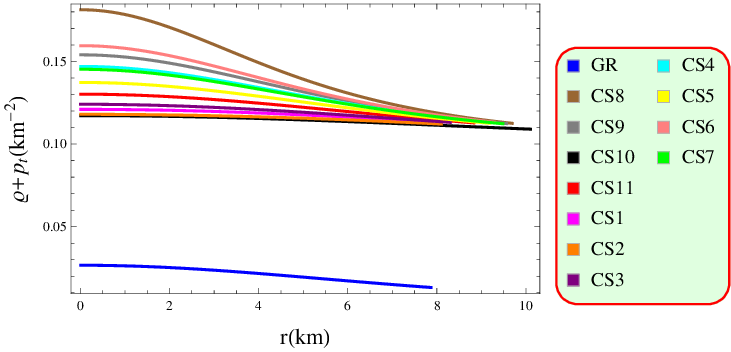,width=.6\linewidth}
\epsfig{file=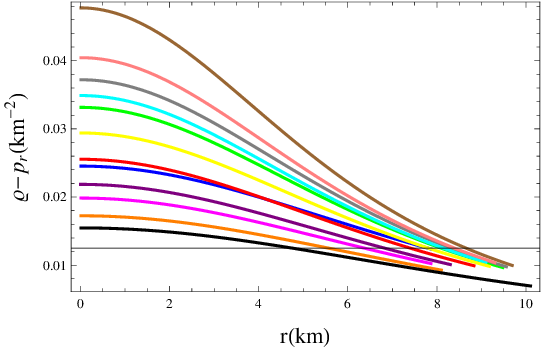,width=.45\linewidth}\epsfig{file=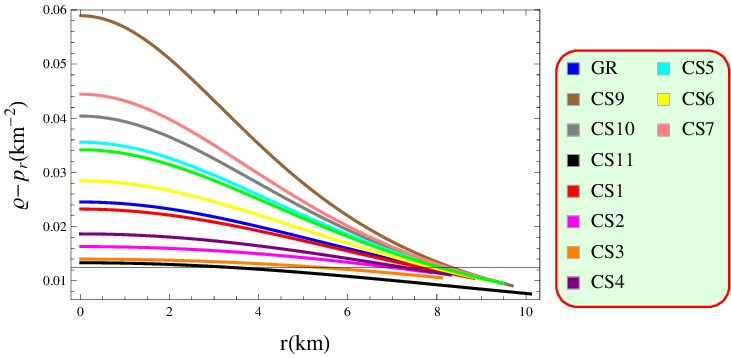,width=.6\linewidth}
\epsfig{file=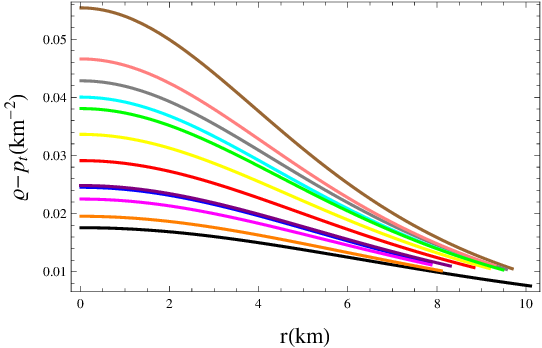,width=.45\linewidth}\epsfig{file=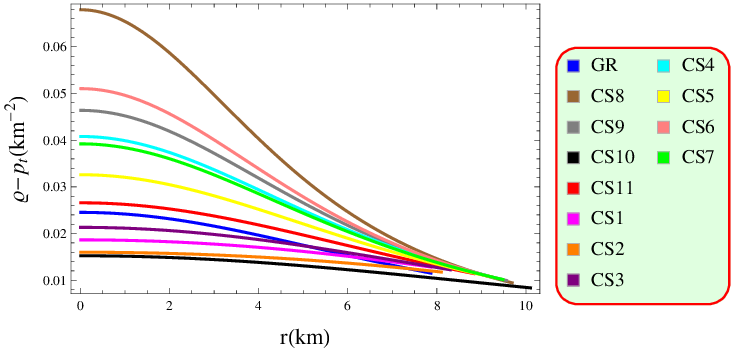,width=.6\linewidth}
\caption{Energy bounds against radial coordinate for solutions
\textbf{I}(left) and \textbf{II}(right) for $\alpha=-0.7$ and
$\beta=-0.1$.}
\end{figure}

\subsection{State Parameters Analysis}
\begin{figure}\center
\epsfig{file=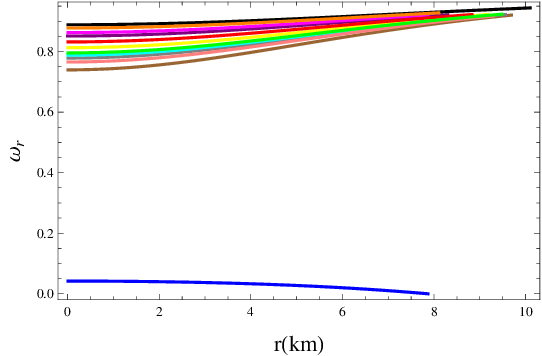,width=.45\linewidth}\epsfig{file=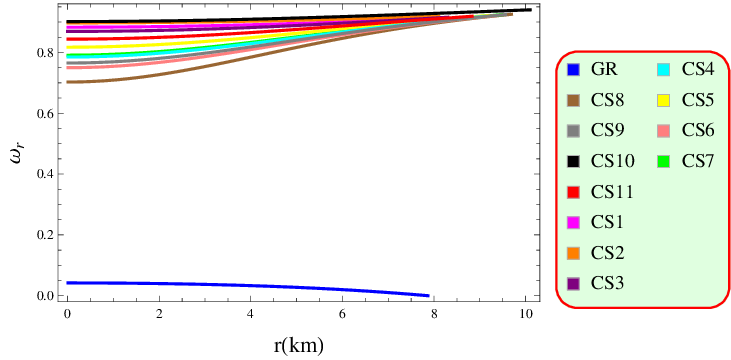,width=.6\linewidth}
\epsfig{file=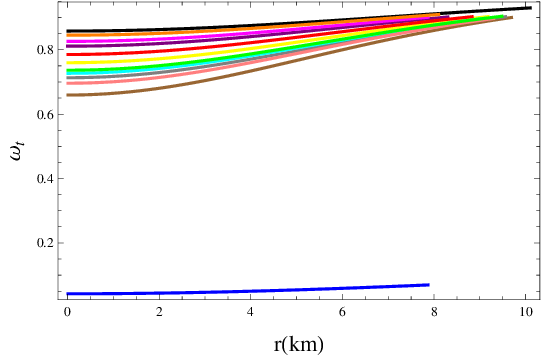,width=.45\linewidth}\epsfig{file=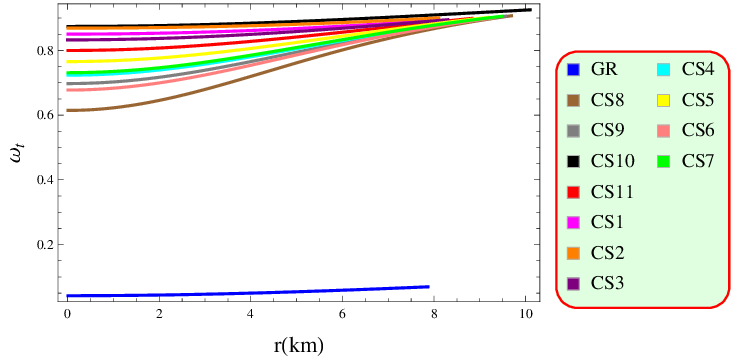,width=.6\linewidth}\caption{The
EoS parameters against radial coordinate for both solutions
\textbf{I}(left) and \textbf{II}(right) for $\alpha=-0.5$ and
$\beta=-0.2$.}
\end{figure}
\begin{figure}\center
\epsfig{file=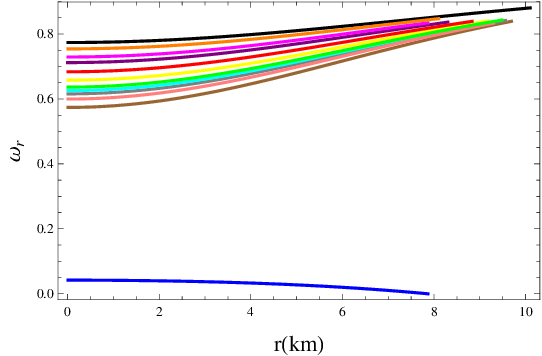,width=.45\linewidth}\epsfig{file=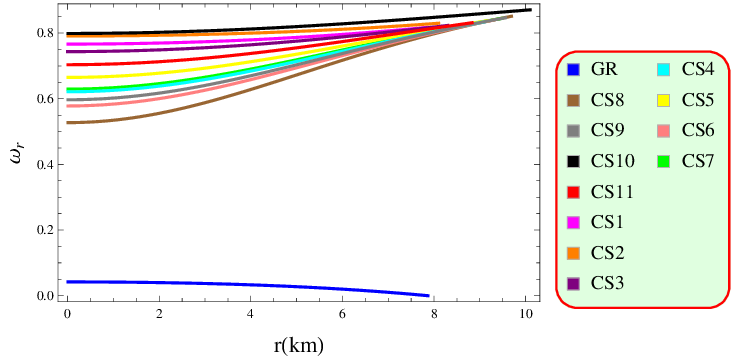,width=.6\linewidth}
\epsfig{file=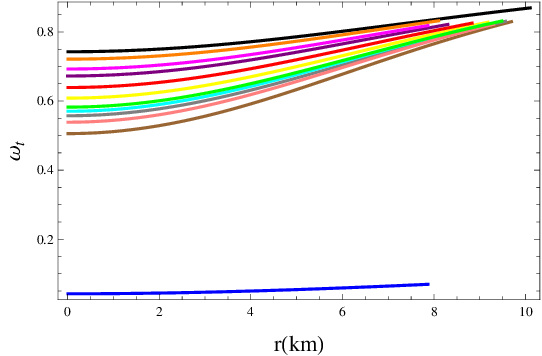,width=.45\linewidth}\epsfig{file=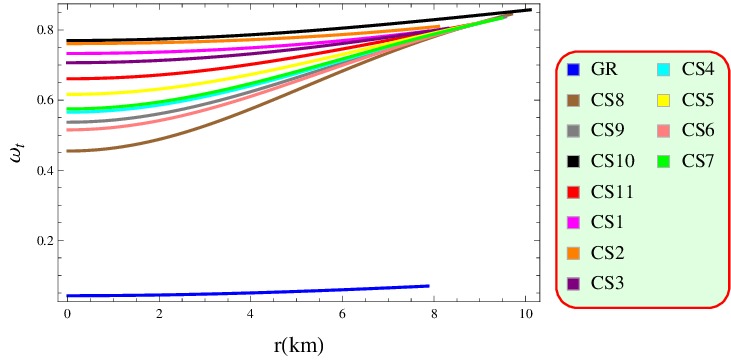,width=.6\linewidth}\caption{The
EoS parameters against radial coordinate for both solutions
\textbf{I}(left) and \textbf{II}(right) for $\alpha=-0.7$ and
$\beta=-0.1$.}
\end{figure}

The equation of state (EoS) is an essential factor that
characterizes the thermodynamic features of a system by connecting
its pressure with density. In the field of astrophysics, this
parameter is significant to understand the internal structure and
stability of CSs. It determines how matter behaves under extreme
conditions, affecting mass-radius relationships. Various forms of
the EoS can be used to model different astrophysical frameworks. The
barotropic is a simple EoS that directly relates a system's pressure
to its energy density as $ \omega_{r}= \frac{p_{r}}{\varrho},~
\omega_{t}=\frac{p_{t}}{\varrho}$. The expressions of EoS parameters
for solutions \textbf{I} and \textbf{II} are given in Appendix
\textbf{C}. The EoS parameters in the range of [0,1] define feasible
CSs \cite{62}. Figures \textbf{9} and \textbf{10} confirm the
physical viability of the stellar structures for both solutions.

\subsection{Study of Some Physical Aspects}

The mass of CSs is determined as
\begin{equation}\nonumber
M=4\pi\int^{\mathbf{R}}_{0} r^{2}\varrho dr
\end{equation}
We use $M(0)=0$ condition to find numerical solution of this
equation. The behavior of mass distribution in CSs with the initial
condition $M(0)=0$ for both solutions is shown in Figures
\textbf{11} and \textbf{12}. The graph shows that the mass function
increases as radius increases and approaches to 0 at the center,
which indicates that there is no singularities in mass distribution.
Several physical parameters help to analyze the viability of
celestial objects. The surface redshift is another important
physical parameter which helps to examine the geometry of CSs. This
measures the gravitational impact on the properties of light emitted
by CSs. Additional information about the object's internal structure
and gravitational field can be gained by expressing this redshift
function as $(Z_s = \frac{1}{\sqrt{1-2\mu}}-1)$. According to a
critical condition set by Buchdahl \cite{62a}, a celestial object
with a perfect matter distribution must have a surface redshift of
less than 2. In contrast, Ivanov \cite{63} found that anisotropic
configurations had a limitation of 5.211 for feasible CSs. The
behavior of redshift functions satisfied viability conditions of CSs
for both solutions as shown in Figures \textbf{11} and \textbf{12}.

\subsection{Mass-Radius and Mass-Inertia Relations}

The mass-radius $(M-\mathbf{R})$ and mass-inertia $(M-I)$ relations
are essential tools in the study of CSs \cite{76}. These provide
insights into the relationship between mass, radius and moment of
inertia for the CSs \cite{77}. The mass-radius relationship helps in
understanding how the mass of a CS influences its size and structure
\cite{78}. The $M-R$ relation helps astronomers to compare
theoretical models with observational data, allowing them to
constrain the properties of CSs and understand their internal
structure. The $M-I$ shows the relationship between the mass and the
moment of inertia for a CS \cite{79}. Moment of inertia is a measure
of an object's resistance to changes in rotation. It depends not
only on the mass distribution but also on the object's shape and
size. In the context of CSs, the $M-I$ relation provides insights
into how mass affects the rotational properties of these CSs. For
example, the CSs are known for their rapid rotation. As the mass of
a CSs increases, its moment of inertia also increases, affecting its
rotational behavior. The $M-I$ relation is very useful for studying
CSs, which are rapidly rotating neutron stars emitting beams of
electromagnetic radiation. By observing the CSs rotation period and
understanding its moment of inertia, astronomers can infer
properties such as its mass and internal structure. Both the
mass-radius and mass-inertia relations are crucial tools for
understanding the properties and behavior of CSs, providing valuable
insights into the nature of these fascinating objects.
\begin{figure}\center
\epsfig{file=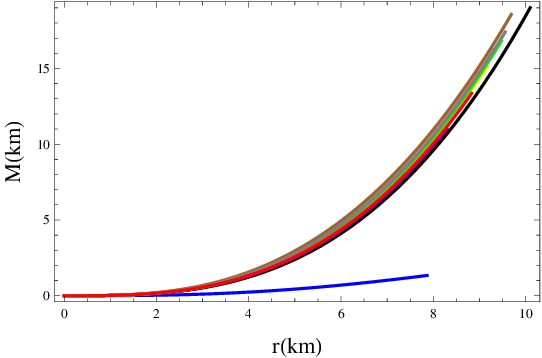,width=.45\linewidth}\epsfig{file=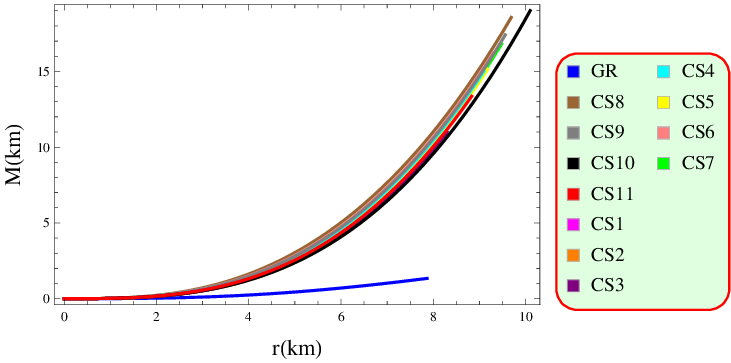,width=.6\linewidth}
\epsfig{file=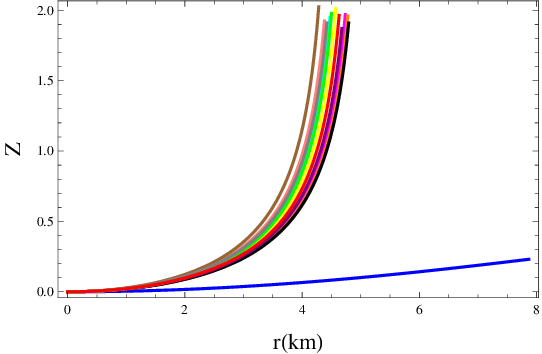,width=.45\linewidth}\epsfig{file=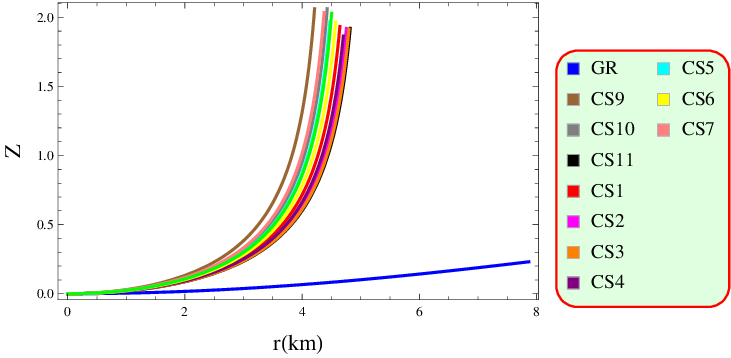,width=.6\linewidth}
\caption{Physical features against radial coordinate for solutions
\textbf{I}(left) and \textbf{II}(right) for $\alpha=-0.5$ and
$\beta=-0.2$.}
\end{figure}
\begin{figure}\center
\epsfig{file=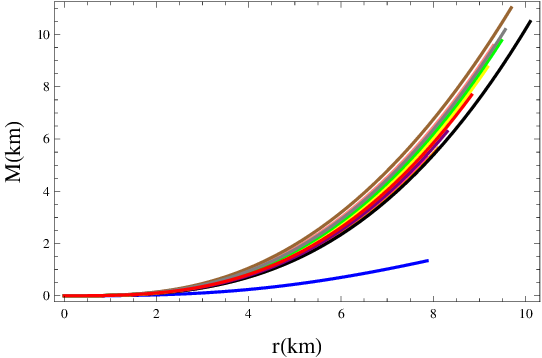,width=.45\linewidth}\epsfig{file=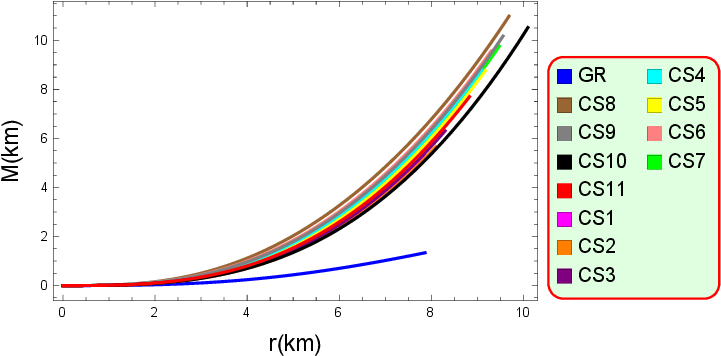,width=.6\linewidth}
\epsfig{file=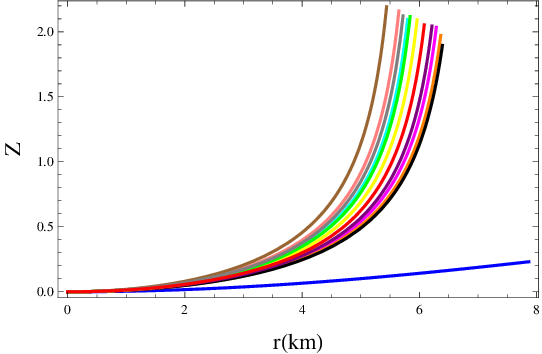,width=.45\linewidth}\epsfig{file=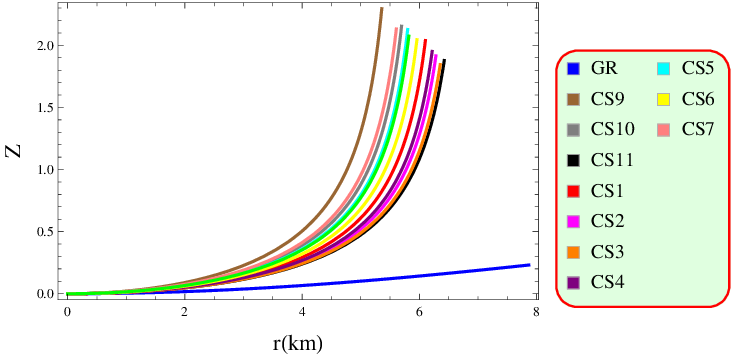,width=.6\linewidth}
\caption{Physical features against radial coordinate for solutions
\textbf{I}(left) and \textbf{II}(right) for $\alpha=-0.7$ and
$\beta=-0.1$.}
\end{figure}

Here, we investigate the gravitational mass and radius with the
following condition $e^{\eta}=1-\frac{2M}{\mathbf{R}}$. The
relationship between the total mass (expressed in $M_{\odot}$) and
the radius (measured in $km$) is presented graphically in Figure
\textbf{13} for both solutions. One notable finding from our
graphical analysis is the identification of the maximum mass
exhibited by a star, which we determine to be 5.0 solar mass. This
particular observation corresponds precisely with existing
literature and is consistent with the value provided in Table
\textbf{1} of our study. Specifically, this maximum mass was
associated with the star identified as EXO 1785-248. Such
compatibility between our findings and established data reinforces
the reliability and validity of our research methodology and
results.
\begin{figure}
\epsfig{file=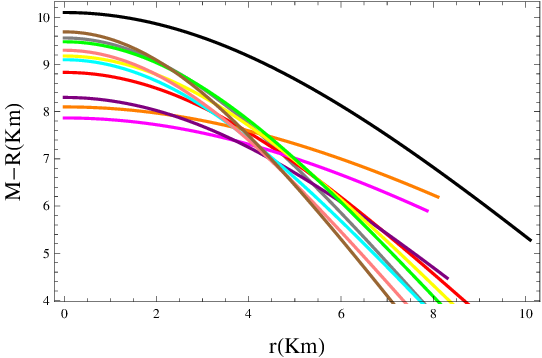,width=.45\linewidth}\epsfig{file=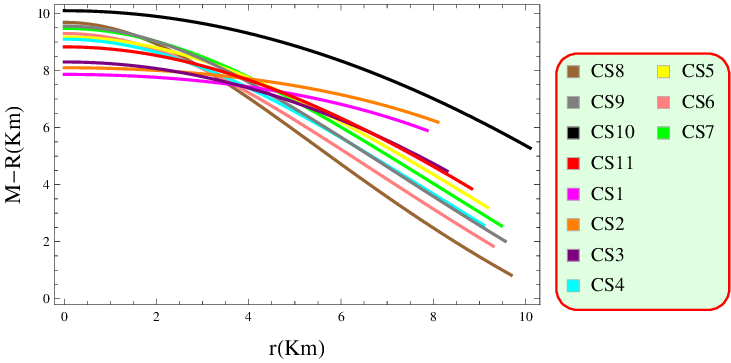,width=.6\linewidth}\caption{M-R
against radial coordinate for solutions \textbf{I}(left) and
\textbf{II}(right).}
\end{figure}
The study of moment of inertia is crucial in understanding the
rotational dynamics of CSs. In the context of astrophysics, the
moment of inertia provides insight into how mass is distributed in
an object and how it affects its rotational behavior. Understanding
the relationship between mass and moment of inertia is essential for
modeling and predicting the behavior of CSs in astrophysical
scenarios, providing valuable insights into their structure and
dynamics. The formula for the moment of inertia as suggested by
Bejger-Haensel
\begin{equation}\label{57b}
I=\frac{2}{5}(1+M/M_{\odot})MR^2,
\end{equation}
allows us to relate the static properties of an object to its
rotational characteristics. The maximum mass of a uniformly slow
rotating configuration provides an approximate estimation of the
moment of inertia. This means that for a given mass and radius, the
moment of inertia reaches its maximum value when the CS is rotating
at its slowest possible rate. Examining the nature of the moment of
inertia with respect to mass, it is evident that as mass increases
the moment of inertia also increases. This relationship is depicted
in Figure \textbf{14} for both solutions. As mass increases, the
more mass is distributed farther away from the axis of rotation,
resulting in a larger moment of inertia. Eventually, the moment of
inertia reaches its maximum value for the given mass, indicating
that further increases in mass do not significantly affect the
rotational dynamics.
\begin{figure}
\epsfig{file=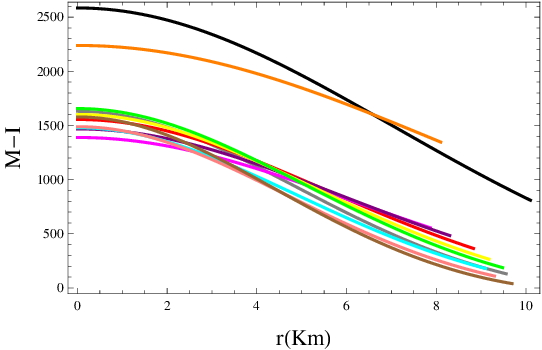,width=.45\linewidth}\epsfig{file=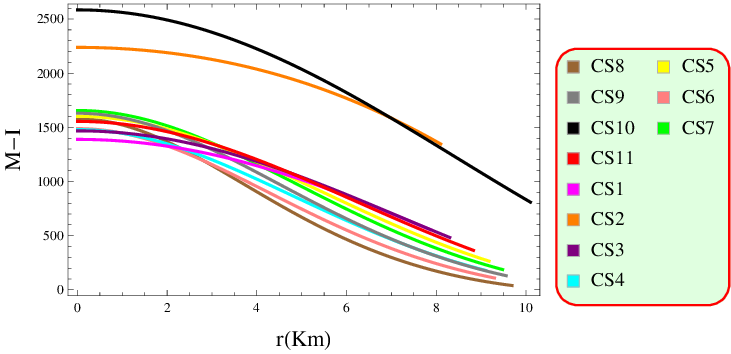,width=.6\linewidth}\caption{M-I
against radial coordinate for solutions \textbf{I}(left) and
\textbf{II}(right).}
\end{figure}

\section{Stability Analysis}

Stability plays a fundamental role in understanding self-gravitating
systems. This method examines matter behavior in astrophysical
objects, evaluating their resistance to gravitational collapse and
assesses the viability and consistency of cosmic configurations.
Here, we check the stability of the considered CSs by causality
condition and Herrera's cracking approach.

\subsection{Causality Constraint}
\begin{figure}
\epsfig{file=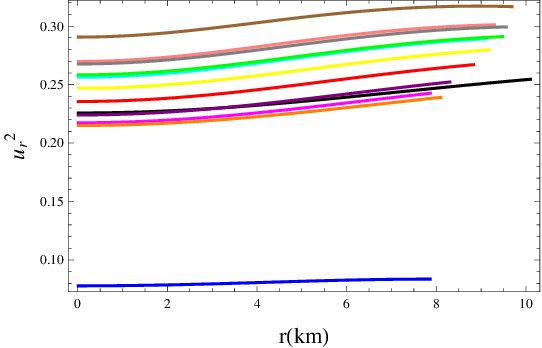,width=.45\linewidth}\epsfig{file=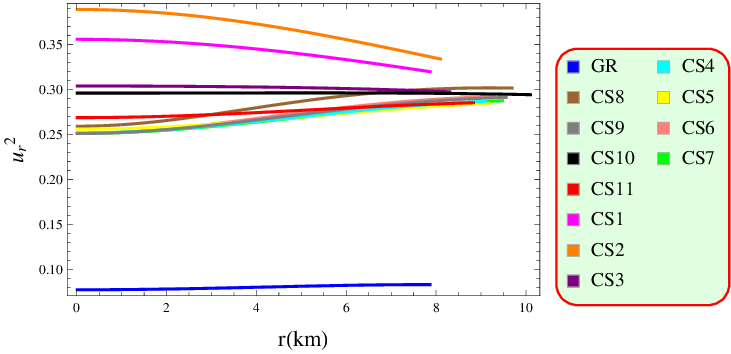,width=.6\linewidth}
\epsfig{file=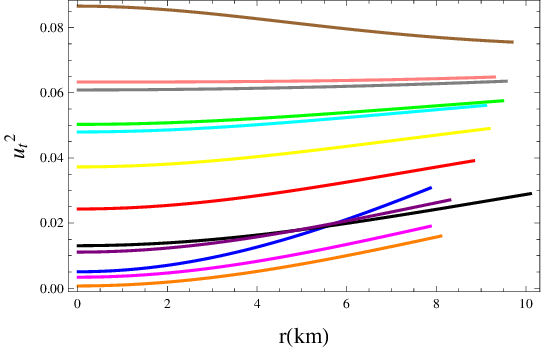,width=.45\linewidth}\epsfig{file=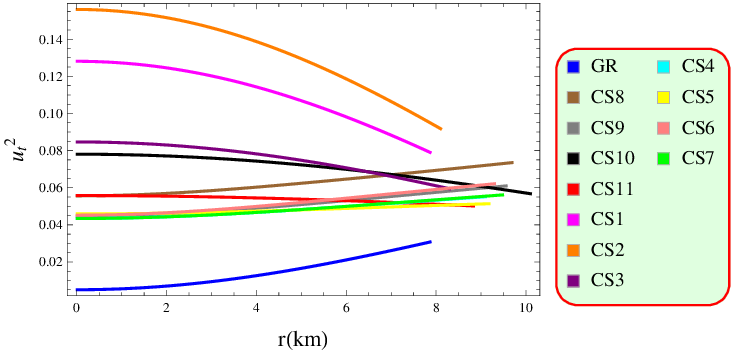,width=.6\linewidth}
\caption{Causality condition against radial coordinate for both
solutions \textbf{I}(left) and \textbf{II}(right) for $\alpha=-0.5$
and $\beta=-0.2$.}
\end{figure}
\begin{figure}
\epsfig{file=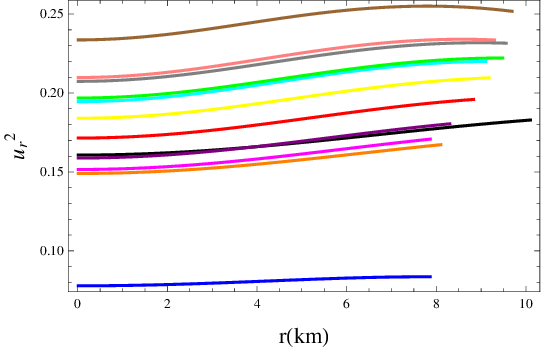,width=.45\linewidth}\epsfig{file=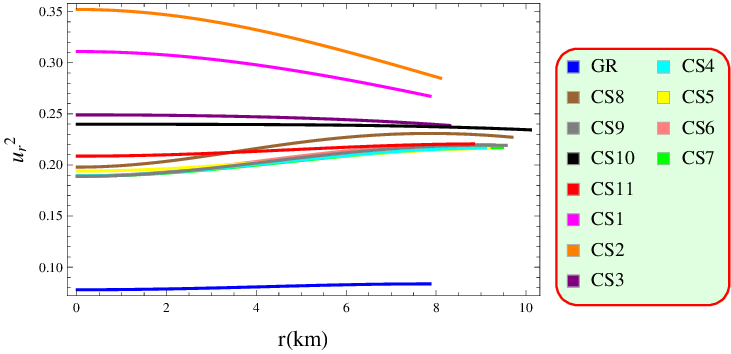,width=.6\linewidth}
\epsfig{file=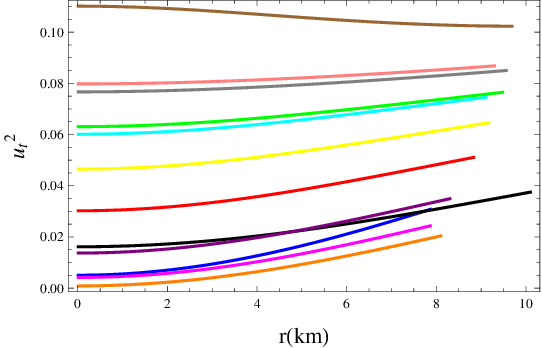,width=.45\linewidth}\epsfig{file=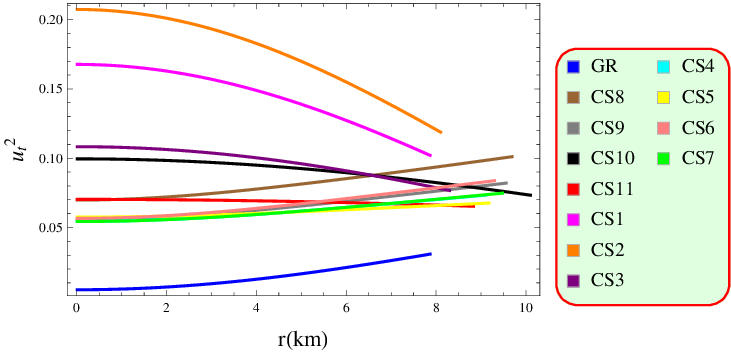,width=.6\linewidth}
\caption{Causality condition against radial coordinate for both
solutions \textbf{I}(left) and \textbf{II}(right) for $\alpha=-0.7$
and $\beta=-0.1$.}
\end{figure}

The causality condition requires that the time-like interval between
any two events in spacetime must remain positive and less then one.
According to causality condition, the radial $(u_{r})^{2}$ and
tangential $(u_{t})^{2}$ components of squared sound speed must lie
in $[0,1]$ for stable CSs \cite{64}. The components of the squared
sound speed are given by $(u_{r})^{2}=\frac{dp_r}{d\varrho},
(u_{t})^{2}=\frac{dp_t}{d\varrho}$. The expressions of the radial
and tangential components of squared sound speed for both solutions
are given in Appendix \textbf{D}. Figures \textbf{15} and
\textbf{16} demonstrate that the stellar interiors are stable in the
presence of modified terms as they meet the required constraints for
both solutions.

\subsection{Herrera Cracking Approach}

The Herrera cracking method serves as an important tool in
astrophysics, providing a framework to examine the stability of
self-gravitating fluid configurations when subjected to
perturbations. As proposed by Herrera, a cosmic structure is
consider to be stable when the difference between $(u_{t}^{2})$ and
$(u_{r}^{2})$ falls in 0 and 1 \cite{64a}. If this condition is
violated, the system may undergo cracking, leading to instability.
Figures \textbf{17} and \textbf{18} indicate that physically
feasible and stable CSs exist in this theory for both solution.
\begin{figure}
\epsfig{file=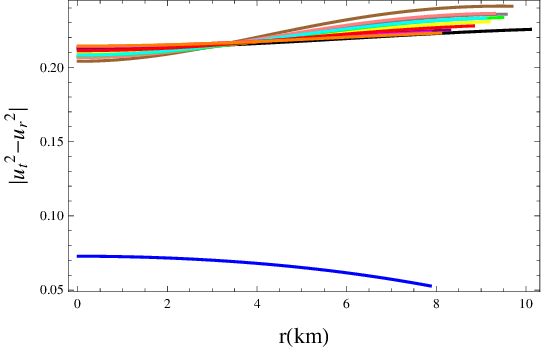,width=.45\linewidth}\epsfig{file=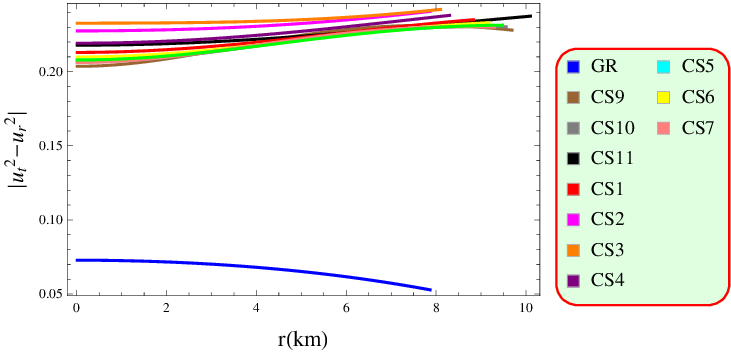,width=.6\linewidth}
\caption{Herrera condition against radial coordinate for both
solutions \textbf{I}(left) and \textbf{II}(right) for $\alpha=-0.5$
and $\beta=-0.2$.}
\end{figure}
\begin{figure}
\epsfig{file=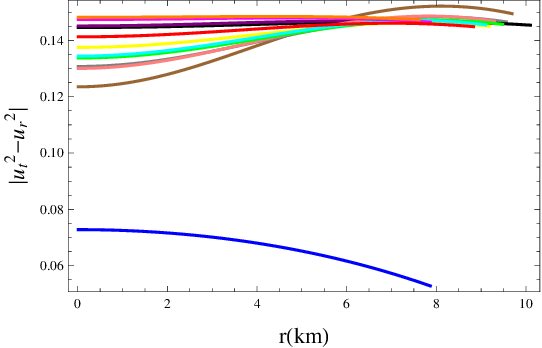,width=.45\linewidth}\epsfig{file=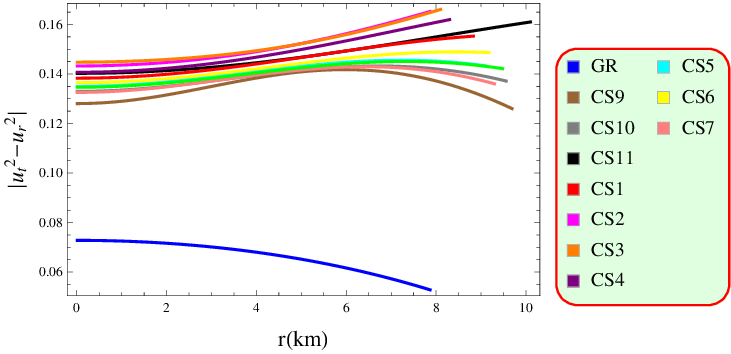,width=.6\linewidth}
\caption{Herrera condition against radial coordinate for both
solutions \textbf{I}(left) and \textbf{II}(right) for $\alpha=-0.7$
and $\beta=-0.1$.}
\end{figure}

\section{Conclusions}

This study examines the viability and stability of stellar interiors
corresponding to two distinct non-singular solutions in the
$f(Q,L_{m})$ theory. The unknown constants in the metric
coefficients are calculated by Darmois junction conditions (Tables
\textbf{2} and \textbf{3} for solutions \textbf{I} and \textbf{II},
respectively). We have used a specific $f(Q,L_{m})$ model to analyze
feasibility and stability of the CSs across various parameters. The
motivation to explore this alternative framework comes form its
crucial theoretical and empirical factors in gravitational physics
and cosmology. The matter term in the functional action is
significant to the dynamical equations, providing novel clarity on
cosmic progression. This proposed theory offers an explanation for
accelerating universe without depending on a cosmological constant,
resolving DE problem. This modified proposal has garnered
significant attention due to its alignment with various
observational and experimental constraints. The connection between
matter term and non-metricity enhances the viability and enables a
broad spectrum of feasible solutions. Consequently, this theoretical
framework opens up the new avenues for developing cosmic frameworks
that align with current observations. The main motivation for
exploring the CSs in this theory is to better understand the cosmic
structures.

The metric coefficients corresponding to both solutions have regular
spacetime by exhibiting a minimum value at the center and increasing
behavior along the boundary (Figures \textbf{1} and \textbf{2}). The
positive and maximum behavior of matter contents at the center
region and reduce towards boundary ensuring the dense profile of CSs
(Figures \textbf{3} and \textbf{5}). The matter variables have
negative matter gradient, indicating their more compactness nature
(Figures \textbf{4} and \textbf{6}). The interior of CSs has
ordinary matter due to positive behavior of energy constraints
(Figures \textbf{7} and \textbf{8}). The EoS parameters satisfy the
required condition of viability (Figures \textbf{9} and
\textbf{10}). The mass function increases with radial distance and
vanish at the core (Figures \textbf{11} and \textbf{12}). The
redshift function remains less than $5.21$ ensuring the existence of
viable CSs in this theory (Figures \textbf{11} and \textbf{12}). The
relation of $M-R$ and $M-I$ ensure the existence of viable CSs in
this theory (Figures \textbf{13} and \textbf{14}).  The stability of
the considered CSs is confirmed (Figures \textbf{15}-\textbf{18}).
We have evaluated multiple important physical parameters that
fulfill criteria, indicating the existence of viable and stable CSs
in this theory.

Notably, we have observed that all parameters attain their maximum
values when compared to those in GR \cite{66}-\cite{68} and other
modified gravitational theories \cite{69}-\cite{71}. Within the
context of $f(R)$ theory, results indicate the instability of the
Her X-1 compact star for the second gravity model, as the physical
quantities only satisfy a limited range of viability \cite{72}. In
$f(G,T)$ theory, analysis shows that the evolution of compact stars
SAX J1808.4-3658 and 4U 1820-30 is consistent with all three gravity
models, whereas for Her X-1, certain restrictions imposed by the
second model must be met for complete physical viability \cite{73}.
It has been observed that the CSs in the $f(R,\mathcal{T}^{2})$
theory exhibit instability at their core \cite{65}. However, our
graphs indicate that CSs are feasible and stable at the center in
this theory for both solutions.

\vspace{0.25cm}

\section*{Appendix A: Calculation of $Q$}

\renewcommand{\theequation}{A\arabic{equation}}
\setcounter{equation}{0}

\begin{eqnarray}\label{A1}
Q&\equiv&-g^{\xi\tau}\big(L^{\gamma}_{~\vartheta\xi}
L^{\vartheta}_{~\tau\gamma}-
L^{\gamma}_{~\vartheta\gamma}L^{\vartheta}_{~\xi\tau}\big),
\\\label{A2}
L^{\gamma}_{~\vartheta\xi}&=&-\frac{1}{2}g^{\gamma\tau}
\big(Q_{\xi\vartheta\tau}+
Q_{\vartheta\tau\xi}-Q_{\tau\vartheta\xi}\big),
\\\label{A3}
L^{\vartheta}_{~\tau\gamma}&=&-\frac{1}{2}g^{\vartheta\psi}
\big(Q_{\gamma\tau\psi}+ Q_{\tau\gamma\psi}-Q_{\psi\tau\gamma}\big),
\\\label{A4}
L^{\gamma}_{~\vartheta\gamma}&=&-\frac{1}{2}g^{\gamma \psi}
\big(Q_{\gamma\vartheta\psi}+ Q_{\vartheta\psi\gamma
}-Q_{\psi\gamma\vartheta}\big),
\\\label{A5}
L^{\vartheta}_{~\xi\tau}&=&-\frac{1}{2}g^{\vartheta\psi}
\big(Q_{\tau\xi\psi}+ Q_{\xi\psi\tau}-Q_{\psi\xi\tau} \big).
\end{eqnarray}
Therefore, we get
\begin{eqnarray}\label{A6}
-g^{\xi\tau}L^{\gamma }_{~\vartheta\xi}L^{\vartheta}_{~\tau\gamma}
&=&-\frac{1}{4}\big(2Q^{\gamma\tau\gamma}
Q_{\gamma\gamma\tau}-Q^{\gamma\tau\gamma} Q_{\gamma\tau\gamma}\big),
\\\label{A7}
g^{\xi\tau}L^{\gamma}_{~\vartheta\gamma}L^{\vartheta}_{~\xi\tau}
&=&\frac{1}{4}g^{\xi\tau}g^{\vartheta\gamma}Q_\vartheta
\big(Q_{\tau\xi\gamma}+Q_{\xi\gamma\tau} -Q_{\gamma\tau\xi}\big),
\\\label{A8}
Q&=&-\frac{1}{4}\big(Q^{\gamma}Q_{\gamma}-Q^{\gamma\tau\xi}
Q_{\gamma\tau\xi} +2Q^{\gamma\tau\xi}Q_{\xi\gamma\tau}
-2Q^{\gamma}\tilde{Q}_{\gamma}\big).
\end{eqnarray}
Applying Eq.(\ref{23a}), we obtain
\begin{eqnarray}\nonumber
\mathcal{P}^{\gamma\xi\tau}&=&\frac{1}{4}\bigg[-Q^{\gamma\xi\tau}
+Q^{\xi\gamma\tau}+Q^{\tau\gamma\xi}
+Q^{\gamma}g^{\xi\tau}-\tilde{Q}^{\gamma} g^{\xi\tau}-
\frac{1}{2}(g^{\gamma\xi}Q^{\tau}
\\\label{A9}
&+&g^{\gamma\tau}Q^{\xi})\bigg],
\\\label{A10}
-Q_{\gamma\xi\tau}\mathcal{P}^{\gamma\xi\tau}
&=&-\frac{1}{4}\big(-Q^{\gamma\xi\tau}
Q_{\gamma\xi\tau}+2Q_{\gamma\xi\tau} Q^{\xi\gamma\tau}
+Q^{\gamma}Q_{\gamma} -2Q_{\gamma}\tilde{Q}^{\gamma}\big) .
\end{eqnarray}

\section*{Appendix B: Variation of $\delta Q$}

\renewcommand{\theequation}{B\arabic{equation}}
\setcounter{equation}{0}

We take non-metricity as
\begin{eqnarray}\label{B1}
Q_{\gamma\xi\tau}&=&\nabla_{\gamma}g_{\xi\tau},
\\\label{B2}
Q^{\gamma}_{~\xi\tau}&=&g^{\gamma\vartheta} Q_{\vartheta\xi\tau}=
g^{\gamma\vartheta}\nabla_{\vartheta}g_{\xi\tau}=\nabla^{\gamma}
g_{\xi\tau},
\\\label{B3}
Q^{~~\xi}_{\gamma~~\tau}&=&g^{\xi\vartheta}
Q_{\gamma\vartheta\tau}=g^{\xi\vartheta}\nabla_{\gamma}
g_{\vartheta\tau}=-g_{\vartheta\tau}\nabla_{\gamma}g^{\xi\vartheta},
\\\label{B4}
Q^{~~\tau}_{\gamma\xi}&=&g^{\tau\vartheta}
Q_{\gamma\xi\vartheta}=g^{\tau\vartheta}\nabla_{\gamma}
g_{\xi\vartheta}=-g_{\xi\vartheta}\nabla_{\gamma}g^{\tau\vartheta},
\\\label{B5}
Q^{\gamma\xi}_{~~\tau}&=&g^{\gamma\vartheta}
g^{\xi\varepsilon}\nabla_{\vartheta}g_{\varepsilon\tau}
=g^{\xi\varepsilon}\nabla^{\gamma}g_{\varepsilon\tau}=-g_{\varepsilon\tau}
\nabla^{\gamma}g^{\xi\varepsilon},
\\\label{B6}
Q^{\gamma~\tau}_{~\xi}&=&g^{\gamma\vartheta}
g^{\tau\varepsilon}\nabla_{\vartheta}g_{\xi\varepsilon}
=g^{\tau\varepsilon}\nabla^{\gamma}g_{\xi\varepsilon}=-g_{\xi\varepsilon}
\nabla^{\gamma}g^{\tau\varepsilon},
\\\label{B7}
Q^{~~\xi\tau}_{\gamma}&=&g^{\xi\varepsilon}g^{\tau
\vartheta}\nabla_{\gamma}g_{\varepsilon\vartheta}
=-g^{\xi\varepsilon}g_{\varepsilon\vartheta}\nabla_{\gamma}g^{\tau\vartheta}
=-\nabla_{\gamma}g^{\xi\tau},
\\\label{B8}
Q^{\gamma\xi\tau}&=&-\nabla^{\gamma}g_{\xi\tau}.
\end{eqnarray}
Using Eqs.(\ref{B6}) and (\ref{B7}), we obtain
\begin{eqnarray}\nonumber
\delta Q&=&-\frac{1}{4}\delta\bigg(-Q ^{\gamma\tau\xi}
Q_{\gamma\tau\xi} +2Q^{\gamma\tau\xi}Q_{\xi\gamma\tau}
-2Q^{\gamma}\tilde{Q}_{\gamma} +Q^{\gamma}Q_{\gamma}\bigg),
\\\nonumber
&=&-\frac{1}{4}\bigg(-\delta Q^{\gamma \tau\xi} Q_{\gamma\tau\xi}
-Q^{\gamma\tau\xi}\delta Q_{\gamma\tau\xi} +2\delta Q^{\gamma
\tau\xi}Q_{\xi\gamma\tau}
\\\nonumber
&+&2Q^{\gamma \tau\xi}\delta Q_{\xi\gamma\tau} -2\delta
Q^{\gamma}\tilde{Q}_{\gamma} -2Q^{\gamma}\delta\tilde{Q}_{\gamma}
+\delta Q^{\gamma}Q_{\gamma} +Q^{\gamma}\delta Q_{\gamma}\bigg),
\\\nonumber
&=&-\frac{1}{4}\bigg[Q_{\gamma\tau\xi} \nabla^{\gamma}\delta
g^{\tau\xi}-Q^{\gamma \tau\xi} \nabla_{\gamma}\delta
g_{\tau\xi}-2Q_{\xi\gamma\tau} \nabla^{\gamma }\delta
g^{\tau\xi}+2Q^{\gamma\tau\xi}\nabla_{\xi}\delta
g_{\gamma\tau}\\\nonumber
&+&2\tilde{Q}_{\gamma}\nabla^{\gamma}g^{\xi\tau}\delta
g_{\xi\tau}+2\tilde{Q}_{\gamma}g_{\xi\tau}\nabla^{\gamma}\delta
g^{\xi\tau}-2Q^{\gamma}\nabla^{\vartheta}\delta
g_{\gamma\vartheta}-Q_{\gamma}\nabla^{\gamma}g^{\xi\tau}\delta
g_{\xi\tau}\\
\label{B9} &-&Q_{\gamma}g_{\xi\tau}\nabla^{\gamma}\delta
g^{\xi\tau}-Q^{\gamma}\nabla_{\gamma}g^{\xi\tau}\delta
g_{\xi\tau}-Q^{\gamma}g_{\xi\tau}\nabla_{\gamma}\delta
g^{\xi\tau}\bigg].
\end{eqnarray}
Here, we use the following equations as
\begin{eqnarray}\label{B10}
\delta g_{\xi\tau}&=&-g_{\xi\gamma }\delta
g^{\gamma\vartheta}g_{\vartheta\tau},
\\\label{B11}
-Q^{\gamma \tau\varepsilon}\nabla_{\gamma}\delta
g_{\tau\varepsilon}&=&-Q^{\gamma
\tau\varepsilon}\nabla_{\gamma}\big(-g_{\tau\vartheta}\delta
g^{\vartheta\vartheta}g_{\vartheta\varepsilon}\big),
\\\nonumber
-2Q^{\varepsilon}\nabla^{\vartheta}\delta
g_{\varepsilon\vartheta}&=&2Q^{\gamma}Q_{\tau\gamma \xi}\delta
g^{\xi\tau}+2Q_{\xi}\tilde{Q}_{\tau}\delta g^{\xi\tau}
\\\label{B11a}
&+&2Q_{\tau}g_{\gamma\varepsilon}\nabla^{\gamma}g^{\tau\varepsilon}.
\end{eqnarray}
Thus, Eq.(\ref{B9}) becomes
\begin{equation}\label{B12}
\delta Q=2\mathcal{P}_{\gamma \tau\varepsilon}\nabla^{\gamma}\delta
g^{\tau\varepsilon}-\big(\mathcal{P}_{\xi\gamma\vartheta}Q^{~\gamma
\vartheta}_{\tau}
-2Q^{\gamma\vartheta}_{~\xi}\mathcal{P}_{\gamma\vartheta\tau}\big)\delta
g^{\xi\tau},
\end{equation}
where
\begin{equation}\label{B13}
2\mathcal{P}_{\gamma\tau\varepsilon}=-\frac{1}{4}\bigg[2Q_{\gamma
\tau\varepsilon}-2Q_{\varepsilon\gamma\tau}-2Q_{\tau\varepsilon\gamma}+2
Q_{\tau}g_{\gamma \varepsilon}+2(\tilde{Q}_{\gamma}-Q_{\gamma})g
_{\tau\varepsilon}\bigg],
\end{equation}
\begin{eqnarray}\nonumber
4\big(\mathcal{P}_{\xi\gamma\vartheta}Q^{~~\gamma \vartheta}_{\tau}
-2Q^{\gamma\vartheta}_{~~\xi}\mathcal{P}_{\gamma\vartheta\tau}\big)&=&
2Q^{\gamma \vartheta}_{~~\tau}Q _{\gamma\vartheta\xi}-4 Q^{~~\gamma
\vartheta}_{\xi}Q_{\vartheta\gamma\tau}
+2\tilde{Q}^{\gamma}Q_{\gamma\xi\tau}
\\\label{B14}
&+&2Q^{\gamma }Q_{\tau\gamma\xi} +2Q_{\xi}\tilde{Q}_{\tau}-
Q^{\gamma}Q_{\gamma\xi\tau}.
\end{eqnarray}

\section*{Appendix C: Calculation of the EoS Parameters for
Solutions \textbf{I} and \textbf{II}}

\renewcommand{\theequation}{C\arabic{equation}}
\setcounter{equation}{0}

Using Eqs.(\ref{16}) and (\ref{18}), we get the expressions for EoS
parameters for solution \textbf{1} as
\begin{eqnarray}\nonumber
\omega_{r}&=&\bigg[(1 + r^2 c) (r^5 s c^2 \beta -8 r s c + 2 a (r^2
c)^{\frac{3}{2}} \beta + r^3 s c (\beta-2 c \alpha) +   2 a
\\\nonumber
&\times& \sqrt{r^2c} (\beta-2 c \alpha))\bigg]\bigg[-16 r s c + r^7
s c^3 \beta + 2 a (r^2 c)^{\frac{5}{2}} \beta + r^3 s c (2 c
(\alpha-8)
\\\label{C1}
&+& \beta) + 2 r^5 s c^2 ( \beta -c \alpha) + 4 a (r^2
c)^{\frac{3}{2}} (\beta -c \alpha) + 2 a \sqrt{r^2c} (2 c \alpha +
\beta)\bigg]^{-1},
\\\nonumber
\omega_{t}&=& \bigg[8 r s c \alpha + 2 r^5 s c^2 \beta+ r^7 s c^3
\beta + 4 a (r^2 c)^{\frac{3}{2}} \beta + 2 a (r^2 c)^{\frac{5}{2}}
\beta + 2 a \sqrt{r^2c} (\beta
\\\nonumber
&-&2 c \alpha) + r^3 s c (2 c \alpha + \beta)\bigg]\bigg[r^7 s c^3
\beta-16 r s c+ 2 a (r^2 c)^{\frac{5}{2}} \beta + r^3 s c (2 c
(\alpha-8)
\\\label{C2}
&+& \beta) + 2 r^5 s c^2 (\beta -c \alpha) + 4 a (r^2
c)^{\frac{3}{2}} (\beta-c \alpha) + 2 a \sqrt{r^2c}(2 c \alpha +
\beta) \bigg]^{-1}.
\end{eqnarray}
Using Eqs.(\ref{49a})-(\ref{51a}), we have the expressions for EoS
parameters for solution \textbf{II} as
\begin{eqnarray}\nonumber
\omega_{r}&=&-\bigg[(\frac{A^2 r^2}{(B r^2+1)^4}+1)(\frac{2
\alpha(\frac{4 A B \sigma r^2}{(B r^2+1)(2 \imath B (B r^2+1)-A
\sigma)}+1)}{\frac{A^2 r^2}{(B r^2+1)^4}+1}-2 \alpha +\beta
r^2)\bigg]
\\\nonumber
&\times&\bigg[r^2 (\beta(-\frac{A^2 r^2}{(B r^2+1)^4}-1)+\frac{2
\alpha  A^3 r^2 }{(B r^2+1)^{10}(2 \imath B (B r^2+1)-A \sigma)}
\\\nonumber
&\times&((-A (3 B r^2-1)(2 \imath B (B r^2+1)^2 (2 A^2 r^2 +3(B r^2
+1)^4)-A \sigma
\\\nonumber
&\times&(2 A^2 r^2 (B r^2 -1)+(5 B r^2-3)(B r^2+1)^4)))(A^2 r^2+(B
r^2+1)^4)^{-1}
\\\label{C3}
&-&2 B \sigma(B r^2+1)^5))\bigg]^{-1},
\\\nonumber
\omega_{t}&=& \bigg[A^5 \beta  \sigma r^4-2 A^4 \imath B \beta  r^4
(B r^2+1)-4 A^2 \imath B(B r^2+1)^4 (B \beta r^4+3 \alpha B r^2
\\\nonumber
&+&\beta  r^2-\alpha)+A \sigma (B r^2+1)^6 (8 \alpha B (B
r^2-1)+\beta(B r^2+1)^2)+2 \sigma (A B r^2
\\\nonumber
&+&A)^3 (B \beta r^4+\alpha  B r^2+\beta r^2-\alpha)-2 \imath B
\beta (B r^2+1)^9\bigg]\bigg[(B r^2+1)^4 (A^2 r^2
\\\nonumber
&+&(B r^2+1)^4)(2 \imath B (B r^2+1)-A \sigma)(\beta(-\frac{A^2
r^2}{(B r^2+1)^4}-1)
\\\nonumber
&+& \frac{2 \alpha A^3 r^2}{(B r^2+1)^{10}(2 \imath B (B r^2+1)-A
\sigma)}(-A (3 B r^2-1)(2 \imath B (B r^2+1)^2
\\\nonumber
&\times&(2 A^2 r^2+3(B r^2+1)^4)-A \sigma(2 A^2 r^2 (B r^2-1)-(5 B
r^2-3)
\\\label{C4}
&\times&(B r^2+1)^4)))(A^2 r^2+(B r^2+1)^4)^{-1}-2 B \sigma (B
r^2+1)^5))\bigg]^{-1}.
\end{eqnarray}

\section*{Appendix D: Calculation of Causality Components for
Solutions \textbf{I} and \textbf{II}}

\renewcommand{\theequation}{D\arabic{equation}}
\setcounter{equation}{0}

Using Eqs.(\ref{16}) and (\ref{18}), we get the field equations
corresponding to the causality parameters along with solution
\textbf{I} as
\begin{eqnarray}\nonumber
u_{r}^{2}&=&-\bigg[(1 + r^2 c) (8 r a s c + 8 s^2 (r^2
c)^{\frac{3}{2}} + 4 r^3 a s c^2 \alpha+ s^2 (r^2 c)^{\frac{5}{2}}
\alpha+ 4 \sqrt{r^2c} (s^2
\\\nonumber
&+& a^2 c \alpha))\bigg]\bigg[ 16 r a s c + 4 r^3 a s c^2 (4 - 3
\alpha) + s^2 (r^2 c)^{\frac{5}{2}} (16 - 3 \alpha) + 4 r^5 a s c^3
\alpha
\\\label{D1}
&+& r^6 s^2 c^3 \sqrt{r^2c} \alpha + 4 \sqrt{r^2c}(2 s^2 - 3 a^2 c
\alpha) + 4 (r^2 c)^{\frac{3}{2}} (6 s^2 + a^2 c \alpha)\bigg]^{-1},
\\\nonumber
u_{t}^{2}&=& -\bigg[2 (6 r^3 s^2 c + r^5 s^2 c^2 + 6 a s
\sqrt{r^2c}-2 a s (r^2 c)^{\frac{3}{2}} + 2 r (s^2 - 2 a^2 c))\alpha
\bigg]
\\\nonumber
&\times&\bigg[ 16 a s \sqrt{r^2c} + 4 a s (r^2 c)^{\frac{3}{2}} (4 -
3 \alpha) + r^5 s^2 c^2 (16 - 3 \alpha) + r^7 s^2 c^3 \alpha+  4 a s
\\\label{D2}
&\times&(r^2 c){\frac{5}{2}} \alpha + 4 r (2 s^2 - 3 a^2 c \alpha)+
4 r^3 c (6 s^2 + a^2 c \alpha)\bigg]^{-1}.
\end{eqnarray}
Using Eqs.(\ref{49a})-(\ref{51a}), we have the field equations of
causality parameters along solution \textbf{II} as
\begin{eqnarray}\nonumber
u_{r}^2&=&\bigg[(A^2 r^2 + (1 + B r^2)^4) (16 \imath B^3 \sigma (1 +
B r^2)^7 +  A^5 \sigma^2 \alpha - 4 A^4 \imath B \sigma (1 + B
\\\nonumber
&\times& r^2) \alpha -  4 A B^2 (1 + B r^2)^5 (\sigma^2 (1 + B r^2)
- 4 \imath^2 B \alpha) - 8 A^2 \imath B^2 \sigma (1 + B r^2)^3
\\\nonumber
&\times& (2 \alpha-1 + B r^2 (1 + 2 \alpha)) + 4 A^3 B (1 + B r^2)^2
(\imath^2 B \alpha + \sigma^2 (\alpha-1 + B r^2
\\\nonumber
&\times& (2 + \alpha))))\bigg]\bigg[32 \imath B^3 \sigma (1 + B
r^2)^{11} + A^7 \sigma^2 r^2 \alpha - 4 A^6 \imath B \sigma r^2 (1 +
B r^2) \alpha
\\\nonumber
&-& 8 A B^2 (1 + B r^2)^8 ((\sigma + B \sigma r^2)^2 + 2 \imath^2 B
(3 - 7 B r^2) \alpha) -   16 A^2 \imath B^2 \sigma (1
\\\nonumber
&+& B r^2)^7 (-1 - 3 \alpha +   B r^2 (-1 + 7 \alpha)) - A^5 (1 + B
r^2)^2 (2 B \sigma^2 r^2 (4 - 7 \alpha)
\\\nonumber
&+& 3 \sigma^2 \alpha + B^2 (-4 \imath^2 r^2 \alpha + \sigma^2 r^4
(7 \alpha -16)))+4 A^4 \imath B \sigma (1 + B r^2)^3 (3 \alpha + B
\\\nonumber
&\times& r^2 (4 - 14 \alpha + B r^2 ( 7 \alpha -4))) + 4 A^3 B (1 +
B r^2)^4 (\imath^2 B (-3 - 7 B r^2 (B
\\\label{D3}
&\times& r^2 -2)) \alpha + (\sigma + B \sigma r^2)^2 (-2 - 3 \alpha
+ B r^2 (2 + 7 \alpha)))\bigg]^{-1},
\\\nonumber
u_{t}^2&=& \bigg[2 (1 + B r^2)^2 (8 \imath B^3 \sigma (-2 + B r^2)
(1 + B r^2)^8 - 2 A^2 \imath B^2 \sigma (1 + B r^2)^4 (11
\\\nonumber
&+& B r^2 (-8 + B r^2)) +  A^5 \sigma^2 (1 + B r^2 (-1 + B r^2)) - 2
A B^2 (1 + B r^2)^6 (\sigma^2
\\\nonumber
&\times&(B r^2 -3) (1 + B r^2) +  4 \imath^2 B (-2 + 3 B r^2)) - 2
A^4 \imath B \sigma (1 + B r^2) (2 + B
\\\nonumber
&\times& r^2 (5 B r^2 -3)) +   A^3 B (1 + B r^2)^2 (4 \imath^2 B (1
+ B r^2 (3 B r^2 -2)) +  \sigma^2 (1
\\\nonumber
&+& B r^2) (7 + B r^2 (-9 + 8 B r^2)))) \alpha\bigg]\bigg[32 \imath
B^3 \sigma (1 +  B r^2)^{11} + A^7 \sigma^2 r^2 \alpha - 4
\\\nonumber
&\times& A^6 \imath B \sigma r^2 (1 + B r^2) \alpha - 8 A B^2 (1 + B
r^2)^8 ((\sigma + B \sigma r^2)^2 +  2 \imath^2 B (3 - 7 B
\\\nonumber
&\times& r^2) \alpha) - 16 A^2 \imath B^2 \sigma (1 + B r^2)^7 (-1 -
3 \alpha + B r^2 (-1 + 7 \alpha)) - A^5 (1 + B
\\\nonumber
&\times& r^2)^2 (2 B \sigma^2 r^2 (4 - 7 \alpha) + 3 \sigma^2 \alpha
+ B^2 (-4 \imath^2 r^2 \alpha + \sigma^2 r^4 (-16 + 7 \alpha))) + 4
\\\nonumber
&\times& A^4 \imath B \sigma (1 + B r^2)^3 (3 \alpha +  B r^2 (4 -
14 \alpha + B r^2 (-4 + 7 \alpha))) + 4 A^3 B (1
\\\nonumber
&+&B r^2)^4 (\imath^2 B (-3 - 7 B r^2 (-2 + B r^2)) \alpha + (\sigma
+ B \sigma r^2)^2 (-2 - 3 \alpha + B
\\\label{D4}
&\times& r^2 (2 + 7 \alpha)))\bigg]^{-1}.
\end{eqnarray}
\\
\textbf{Data Availability Statement:} No new data were generated or
analyzed in support of this research.

\end{document}